\documentclass[aps,prx,twocolumn,superscriptaddress,notitlepage,amsmath,amssymb,citeautoscript,longbibliography,floatfix,10pt]{revtex4-2}
\usepackage[T1]{fontenc}
\usepackage[utf8]{inputenc}
\usepackage{hyperref}
\usepackage{amssymb}
\usepackage{amsmath}
\usepackage{mathtools}
\usepackage{braket}
\usepackage{color}
\usepackage[amsmath]{empheq}
\usepackage{tikz}
\usepackage{color}
\usepackage{nicefrac}
\usepackage{bbm}
\usepackage{tensor}
\usepackage{tikzsymbols}
\usepackage{graphicx}
\usepackage{tikz-cd}
\usepackage{natbib}
\usepackage{comment}
\usepackage{amsthm}

\hypersetup{colorlinks,linkcolor=teal,citecolor=blue}
\newcommand\bigzero{\makebox(0,0){\text{\huge0}}}
\newcommand\bigasterisk{\makebox(0,0){\huge*}}

\newcommand\ua{{\uparrow}}
\newcommand\da{{\downarrow}}

\theoremstyle{definition}
\newtheorem{definition}{Example}

\begin{document}
\title{
Quantum many-body spin ratchets
}
\author{Lenart Zadnik}
\email{lenart.zadnik@fmf.uni-lj.si}
\affiliation{Department of Physics, Faculty of Mathematics and Physics,
University of Ljubljana, Jadranska 19, SI-1000 Ljubljana, Slovenia}
\author{Marko Ljubotina}
\affiliation{Institute of Science and Technology Austria, Am Campus 1, 3400 Klosterneuburg, Austria}
\author{\v{Z}iga Krajnik}
\affiliation{Department of Physics, New York University, 726 Broadway, New York, NY 10003, USA}
\author{Enej Ilievski}
\affiliation{Department of Physics, Faculty of Mathematics and Physics,
University of Ljubljana, Jadranska 19, SI-1000 Ljubljana, Slovenia}
\author{Toma\v{z} Prosen}
\affiliation{Department of Physics, Faculty of Mathematics and Physics,
University of Ljubljana, Jadranska 19, SI-1000 Ljubljana, Slovenia}
\affiliation{Institute of Mathematics, Physics and Mechanics, Jadranska 19, SI-1000, Ljubljana, Slovenia}

\begin{abstract}
Introducing a class of SU(2) invariant quantum unitary circuits generating chiral transport, we examine the role of broken space-reflection and time-reversal symmetries on spin transport properties. Upon adjusting parameters of local unitary gates, the dynamics can be either chaotic or integrable. The latter corresponds to a generalization of the space-time discretized (Trotterized) higher-spin
quantum Heisenberg chain. We demonstrate that breaking of space-reflection symmetry results in a drift in the dynamical spin susceptibility. Remarkably, we find a universal drift velocity given by a simple formula which, at zero average magnetization, depends only on the values of SU(2) Casimir invariants associated with local spins. In the integrable case, the drift velocity formula is confirmed analytically based on the exact solution of thermodynamic Bethe ansatz equations. 
Finally, by inspecting the large fluctuations of the time-integrated current between two halves of the system in stationary maximum-entropy states, we demonstrate violation of the Gallavotti--Cohen symmetry, implying that such states cannot be regarded as equilibrium ones. We show that the scaled cumulant generating function of the time-integrated current instead obeys a generalized fluctuation relation.

\end{abstract}
\maketitle

\section{Introduction}

Recent rapid advancements of quantum computing platforms based on trapped ions, ultracold atoms, and superconducting qubits~\cite{lanyon2011,barreiro2011,blatt2012,salathe2015,barends2015,bernien2017,gross2017,arute2019,carusotto2020,kjaergaard2020,monroe2021,ebadi2021,bravyi2022,daley2022} have drawn considerable attention to dynamics of unitary quantum  circuits and cellular automata~\cite{arrighi2019,farrelly2020,fisher2023}. Besides allowing for classical simulation~\cite{vidal2004,schollwock2011} which can often be efficient, such discrete space-time dynamical systems can likewise be realized on modern experimental quantum platforms. Moreover, they are often amenable to exact solutions~\cite{gritsev2017,vanicat2018,alba2019,klobas2021,claeys2022,gombor2022,miao2023,fritzsch2023,yu2024} and therefore prove particularly useful for benchmarking quantum devices~\cite{aleiner2021,maruyoshi2023,zhang2024}. On the other hand, quantum circuits can exhibit a diverse range of unconventional dynamical properties which include anomalous transport~\cite{bertini2021,bulchandani2021}, recently observed in experiments~\cite{wei2022,joshi2022}, and robustness to integrability breaking perturbations~\cite{surace2024,hudomal2024,surace2023,orlov2023,kurlov2022}. In this view, they are of fundamental theoretical interest to the statistical physics community.

The study of anomalous transport has been at the forefront of theoretical interest in the recent years. Among the most emblematic examples is the discovery~\cite{znidaric2011} of universal superdiffusive transport of Noether charges in integrable models with nonabelian symmetries~\cite{ilievski2021}. The precise determination of the dynamical universality class remains an open question: despite the dynamical two-point function of the charge density coinciding with the scaling function of the Kardar--Parisi--Zhang equation (KPZ) at late times~\cite{ljubotina2017,ilievski2018,ljubotina2019-2,denardis2019,dupont2020,bulchandani2020,bulchandani2020-2,ilievski2021,bulchandani2021,ye2022}, it has been shown that the full probability distribution of net charge transfer is not compatible with the behavior of fluctuations predicted by the KPZ equation~\cite{krajnik2024}. This discrepancy has also been supported by a recent experiment using superconducting quantum processors~\cite{rosenberg2024}.

Reliable extraction of transport coefficients and 
statistical properties of macroscopic fluctuating observables is in practice hindered by the  complexity of simulating strongly interacting quantum dynamics on classical computers. This difficulty can fortunately be overcome in integrable models, where the underlying quasiparticle structure often permits derivation of exact results valid on hydrodynamic scales. Central to this endeavour are the tools of generalized hydrodynamics (GHD)~\cite{alvaredo2016,bertini2016} and ballistic (macroscopic) fluctuation theory~\cite{myers2020,doyon2023}.

\begin{figure}[ht!]
    \centering
    \begin{tikzpicture}
    \draw[black, opacity=0.075, line width=0.2mm] (-4+0.075,0-0.075) -- (0+0.075,4-0.075);
    \draw[black, opacity=0.075, line width=0.2mm] (-4+0.075,2-0.075) -- (-2+0.075,4-0.075);
    \draw[black, opacity=0.075, line width=0.2mm] (-2+0.075,0-0.075) -- (2+0.075,4-0.075);
    \draw[black, opacity=0.075, line width=0.2mm] (0+0.075,0-0.075) -- (4+0.075,4-0.075);
    \draw[black, opacity=0.075, line width=0.2mm] (2+0.075,0-0.075) -- (4+0.075,2-0.075);
    \draw[black, opacity=0.075, line width=0.6mm] (-3+0.075,0-0.075) -- (-4+0.075,1-0.075);
    \draw[black, opacity=0.075, line width=0.6mm] (-1+0.075,0-0.075) -- (-4+0.075,3-0.075);
    \draw[black, opacity=0.075, line width=0.6mm] (1+0.075,0-0.075) -- (-3+0.075,4-0.075);
    \draw[black, opacity=0.075, line width=0.6mm] (3+0.075,0-0.075) -- (-1+0.075,4-0.075);
    \draw[black, opacity=0.075, line width=0.6mm] (4+0.075,1-0.075) -- (1+0.075,4-0.075);
    \draw[black, opacity=0.075, line width=0.6mm] (4+0.075,3-0.075) -- (3+0.075,4-0.075);
    \foreach \x in {0,...,3}
    {
    \fill[white!90!black, opacity=1, rounded corners = 2] (-3.75+2*\x+0.075,0.25-0.075) rectangle ++(0.5,0.5);
    \fill[white!90!black, opacity=1, rounded corners = 2] (-2.75+2*\x+0.075,1.25-0.075) rectangle ++(0.5,0.5);
    \fill[white!90!black, opacity=1, rounded corners = 2] (-3.75+2*\x+0.075,2.25-0.075) rectangle ++(0.5,0.5);
    \fill[white!90!black, opacity=1, rounded corners = 2] (-2.75+2*\x+0.075,3.25-0.075) rectangle ++(0.5,0.5);
    }
    \draw[black,line width=0.2mm] (-4,0) -- (0,4);
    \draw[black,line width=0.2mm] (-4,2) -- (-2,4);
    \draw[black,line width=0.2mm] (-2,0) -- (2,4);
    \draw[black,line width=0.2mm] (0,0) -- (4,4);
    \draw[black,line width=0.2mm] (2,0) -- (4,2);
    \draw[blue!60!white,line width=0.6mm] (-3,0) -- (-4,1);
    \draw[blue!60!white,line width=0.6mm] (-1,0) -- (-4,3);
    \draw[blue!60!white,line width=0.6mm] (1,0) -- (-3,4);
    \draw[blue!60!white,line width=0.6mm] (3,0) -- (-1,4);
    \draw[blue!60!white,line width=0.6mm] (4,1) -- (1,4);
    \draw[blue!60!white,line width=0.6mm] (4,3) -- (3,4);
    \foreach \x in {0,...,3}
    {
    \draw[fill=white, opacity=1, rounded corners = 2,thick] (-3.75+2*\x,0.25) rectangle ++(0.5,0.5);
    \draw[fill=white, opacity=1, rounded corners = 2,thick] (-2.75+2*\x,1.25) rectangle ++(0.5,0.5);
    \draw[fill=white, opacity=1, rounded corners = 2,thick] (-3.75+2*\x,2.25) rectangle ++(0.5,0.5);
    \draw[fill=white, opacity=1, rounded corners = 2,thick] (-2.75+2*\x,3.25) rectangle ++(0.5,0.5);
    }
    \foreach \x in {0,...,3}
    {
    \draw[fill=red!10!green!70!blue!20, opacity=1, rounded corners = 2,thick] (-3.75+2*\x,0.25) rectangle ++(0.5,0.5);
    \draw[fill=red!10!green!70!blue!20, opacity=1, rounded corners = 2,thick] (-2.75+2*\x,1.25) rectangle ++(0.5,0.5);
    \draw[fill=red!10!green!70!blue!20, opacity=1, rounded corners = 2,thick] (-3.75+2*\x,2.25) rectangle ++(0.5,0.5);
    \draw[fill=red!10!green!70!blue!20, opacity=1, rounded corners = 2,thick] (-2.75+2*\x,3.25) rectangle ++(0.5,0.5);
    }
    \foreach \x in {0,...,3}
    {
    \node[anchor=north] at (-4+2*\x,0) {$s_1$};
    \node[anchor=north] at (-3+2*\x+0.1,0) {$s_2$};
    \node[anchor=center] at (-3.5+2*\x,0.5) {$U$};
    \node[anchor=center] at (-3.5+2*\x,2.5) {$U$};
    \node[anchor=center] at (-2.5+2*\x,1.5) {$U$};
    \node[anchor=center] at (-2.5+2*\x,3.5) {$U$};
    }
    \end{tikzpicture}
    \caption{A brickwork configuration of quantum unitaries $U$ [see Eq.~\eqref{eq:quantum_gate}], representing a quantum many-body spin ratchet with time flowing upwards. Each two-body local unitary gate involves a permutation and thus swaps the adjacent spin spaces, yielding chiral dynamics of spin species: spins $s_1$ propagate east (i.e., towards the right) and spins $s_2$ west (i.e., towards the left). The (Floquet) propagator $\mathbb{U}$ given by Eq.~\eqref{eq:propagator} corresponds to two layers of the circuit.}
    \label{fig:ratchet_diagram}
\end{figure}
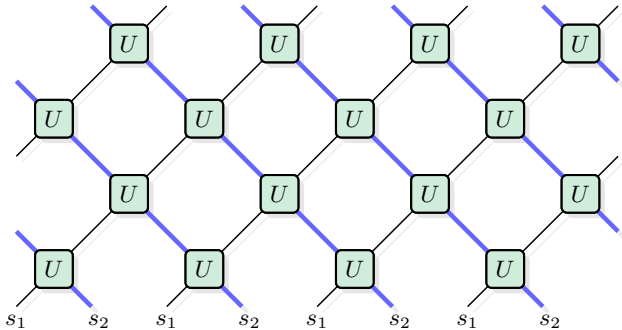

The studies so far have largely investigated anomalous properties of spin or charge transport in interacting many-body systems with unbroken space-time symmetry, i.e., with microscopic dynamics invariant under the time-reversal ($\mathcal{T}$) and space-reflection ($\mathcal{P}$) symmetries.
Instead, we aim to systematically examine the properties of intrinsically chiral microscopic dynamics, in which both $\mathcal{P}$ and $\mathcal{T}$ symmetries are explicitly broken. Such a dynamics is prototypical of quantum ratchets~\cite{reimann1997,denisov2007,salger2009,hamamoto2019}. Our goal here is to devise a many-body analogue of a ratchet, and to investigate how the absence of $\mathcal{P}$ and $\mathcal{T}$ symmetries impacts the charge transport. To this end, we introduce a class of unitary circuits with a brickwork design in the form of a staggered lattice consisting of two alternating spins $s_1$ and $s_2$, schematically presented in Fig.~\ref{fig:ratchet_diagram}. Crucially, in quantum ratchet circuits under consideration the space-reflection symmetry is broken at the level of local unitary gates, rather than by the initial conditions~\cite{bernard2012,bernard2015,bertini2016,alvaredo2016,deluca2017,bertini2018,mazza2018,bertini2019,gruber2019} or by transport-inducing nonunitary boundary processes~\cite{michel2005,michel2008,prosen2009,znidaric2011,mendoza2015,znidaric2016,popkov2020,popkov2020-2} (see also reviews~\cite{prosen2015,bertini2021} and references therein). More specifically, the circuits consist of two-site unitary gates which lack the space-reflection symmetry as a direct consequence of the nearest-neighbor spin exchange.
Mostly for reasons of simplicity, we devote this work to many-body quantum ratchets built out of rotationally symmetric (i.e., SU(2)-invariant) local unitary gates, including both generic (i.e., ergodic) dynamics and exactly solvable (i.e., integrable) instances. As an application, we then characterize spin transport on the ballistic (Euler) scale at a finite magnetization density.

As a consequence of broken space-reflection and time-reversal symmetries, quantum many-body spin ratchets display two notable universal features.
Most remarkably, we demonstrate that
\begin{quote}
    \emph{the dynamical spin susceptibility exhibits a universal non-zero drift velocity, depending only on the size of spins and average background magnetization density.}
\end{quote}
We support this statement by deriving a closed-form expression for the drift velocity using GHD in the integrable ratchets, and by extensive numerical simulations in the nonintegrable ones. In addition, we demonstrate that the drift velocity arises purely from the spin-exchange part of the local unitary gate, enabling us to obtain a general analytic form as a function of both spins and chemical potential.
By numerically studying the spreading of spin fluctuations in the co-moving drift frame, we generically find the anticipated superdiffusive
scaling with dynamical exponent $z=3/2$, as observed also in several other models which possess continuous nonabelian symmetries, and which exhibit dynamical criticality~\cite{znidaric2011,ljubotina2017,ilievski2018,ljubotina2019-2,denardis2019,dupont2020,bulchandani2020,bulchandani2020-2,ilievski2021,bulchandani2021,ye2022,Das2019,Krajnik2020,Evers2020,krajnik2024}.

Our second main result concerns the breaking of time-reversal symmetry. We find that
\begin{quote}
    \emph{in quantum many-body spin ratchets, the Gallavotti--Cohen relation associated with macroscopic current fluctuations is violated in Gibbs states.}
\end{quote}
While we rigorously demonstrate this violation only in the integrable unitary circuits, we conjecture it to remain a general feature of nonintegrable many-body spin ratchets as well. In the integrable ratchet specifically, we compute the third scaled cumulant associated with the time-integrated current density using the results of ballistic (macroscopic) fluctuation theory~\cite{myers2020,doyon2023}. In Gibbs states away from half-filling, we obtain a non-zero value.
Distinctly to time-reversal invariant dynamical systems which always obey the Gallavotti--Cohen relation~\cite{bernard2013,doyon2023}, this fluctuation symmetry is no longer satisfied in quantum spin ratchets. To put it simply, the probability for measuring a large value of current depends on the direction of the flow. This consequently means that Gibbs states are in fact not equilibrium states of quantum ratchets. We instead deduce a generalized fluctuation relation, connecting large current fluctuations in one direction to fluctuations in the opposite direction in a spatially-reflected system.

The remainder of the paper is structured as follows. In Sec.~\ref{sec:ratchets} we introduce quantum many-body ratchets. This is followed by Sec.~\ref{sec:integrability}, where we introduce an integrable quantum ratchet and describe the associated integrable structure. Having discussed the setup, we focus on dynamical properties of the model, which are presented in Sec.~\ref{sec:hydro}. First, in Sec.~\ref{sec:continuity_equations} we define the relevant observables and related continuity equations. Second, in Sec.~\ref{sec:drift_velocity} we compute the first moment of the dynamical structure factor and its drift velocity. Furthermore, we obtain the same expression from a permuting unitary circuit and verify it against tensor network simulations. Lastly, in Sec.~\ref{sec:fluctuations} we discuss large scale fluctuations as an alternative probe of transport. We conclude with a discussion of our results and open questions in Sec.~\ref{sec:discussion}. Several appendices at the end present the details of the calculations.

\section{Quantum spin ratchet circuits}
\label{sec:ratchets}

We consider an inhomogeneous quantum spin chain of length $L\in 2\mathbb{N}$, made out of two spins of not necessarily equal sizes $s_1$ and $s_2$ ($s_1,s_2\in \mathbb{N}/2$), arranged in an alternating fashion. With each spin we associate a local Hilbert space $\mathcal{H}_{s_i}\cong \mathbb{C}^{2s_i+1}$, $i\in\{1,2\}$.

We study a discrete-time two-step unitary evolution of quantum states
\begin{align}
\label{eq:propagator}
     \ket{\psi(t+1)}=\mathbb{U}\ket{\psi(t)},\qquad
     \mathbb{U}=\mathbb{U}_{\rm e}\mathbb{U}_{\rm o},
\end{align}
where $\mathbb{U}_{\rm e}$ and $\mathbb{U}_{\rm o}$ denote
the even-step and odd-step propagators and $t\in\mathbb{N}$ is time. The one-step propagators are composed from local two-site unitary maps $U:\mathcal{H}_{s_1}\otimes\mathcal{H}_{s_2}\to \mathcal{H}_{s_2}\otimes\mathcal{H}_{s_1}$, each acting on two adjacent lattice sites, namely
\begin{align}
    \label{eq:two-steps}
    \mathbb{U}_{\rm e}=\prod_{\ell=1}^{L/2} U_{2\ell,2\ell+1},\qquad \mathbb{U}_{\rm o}=\prod_{\ell=1}^{L/2} U_{2\ell-1,2\ell},
\end{align}
where the subscript indices refer to the the pair of sites on which $U$ acts nontrivially, and periodic boundary conditions have been adopted by identifying $1\equiv L\!+\!1$. Combined together, the unitary maps are arranged in a brickwork architecture as shown in Fig.~\ref{fig:ratchet_diagram}.

We are particularly interested in local quantum unitary maps of the form
\begin{align}
\label{eq:quantum_gate}
    U=P^{s_1,s_2}V,    
\end{align}
where $P^{s_1,s_2}=\sum_{n_1=-s_1}^{s_1}\sum_{n_2=-s_2}^{s_2}\ket{n_2 n_1}\!\bra{n_1  n_2}$ denotes the permutation of two spins, while $V\in{\rm End}(\mathcal{H}_{s_1}\!\otimes\!\mathcal{H}_{s_2})$ is an arbitrary  unitary gate which preserves the ordering of the local degrees of freedom, and which may even differ between the pairs of sites. Such circuits can be viewed as \emph{quantum many-body spin ratchets}~\cite{reimann1997,denisov2007,salger2009,hamamoto2019}: due to permutations, different species of spin get propagated in opposite directions, resulting in a dynamics that breaks space-reflection symmetry. As described in Appendix~\ref{app:embedding}, such a dynamics can be experimentally realized using local quantum unitary gates acting on several copies of identical qubits or qudits. For example, in the case $s_1=1$ and $s_2=1/2$, one can realize $U$ as a quantum unitary gate acting on three qubits (or spins $1/2$).

\subsection{Isotropic spin ratchets}

In this paper we specialize to quantum ratchets composed of SU(2) symmetric gates. This choice is primarily motivated by the recent discovery of anomalous transport properties in integrable models invariant under a continuous nonabelian symmetry
~\cite{ilievski2018,ilievski2021,bulchandani2021}. 
All of the models explicitly considered so far, however, exhibit both space-reflection ($\mathcal{P}$) and time-reversal ($\mathcal{T}$) symmetry.
Our aim is thus to investigate whether absence of $\mathcal{P}$ and $\mathcal{T}$ symmetries has any profound effect on spin transport properties. Additionally, there exist a class of integrable SU(2) symmetric  quantum circuits~\cite{vanicat2018,claeys2022} (and generalizations thereof to other symmetries~\cite{ljubotina2019,medenjak2020,vernier2023,hutsalyuk2024}) which, owing to their particularly simple structure, permit analytical calculations of correlation functions, transport coefficients, and quench dynamics. As outlined below, we will focus on a simple one-parameter family of gates $V$ that, depending on the choice of parameters, encompasses both integrable and ergodic circuits. As such, it is particularly convenient for examining the effects of integrability breaking.

Specifically, global SU(2) symmetry is ensured by realizing unitary gates $V$ in terms of an operator-valued function of the spin magnitude $J\in{\rm End}(\mathcal{H}_{s_1}\!\otimes\!\mathcal{H}_{s_2})$, which is defined through the Casimir invariant
\begin{align}
\label{eq:Casimir_operator}
    J(J\!+\!\mathbbm{1})=(\boldsymbol{S}_1+\boldsymbol{S}_2)^2,\quad \boldsymbol{S}=(S^x,S^y,S^z).
\end{align}
Note that the eigenvalues $j \in \{|s_{1}-s_{2}|,\ldots,s_{1}+s_{2}\}$ of operator $J$ determine the dimensions $2j+1$ of the irreducible components (spin multiplets) in the Clebsch--Gordan decomposition of $\mathcal{H}_{s_1}\otimes\mathcal{H}_{s_2}$. To enable analytical calculations, we choose a one-parameter family of unitary gates $V=R^{s_1,s_2}(\lambda)$ with $\lambda\in\mathbb{R}$, where
\begin{align}
\label{eq:R_matrix}
    R^{s_1\!,s_2}\!(\lambda)\!=\!(\!-\!1)^{J+j_{\rm max}}\!\frac{\Gamma(j_{\rm max}\!+\!1\!+\!i\lambda\!)\Gamma(\!J\!+\![1\!-\!i\lambda]\mathbbm{1}\!)}{\Gamma(j_{\rm max}\!+\!1\!-\!i\lambda\!)\Gamma(\!J\!+\![1\!+\!i\lambda]\mathbbm{1}\!)}\!
\end{align}
obeys the Yang-Baxter equation discussed in Section~\ref{sec:integrability}. Here, $\Gamma$ denotes the Euler $\Gamma$-function and $j_{\rm max}=s_1+s_2$ is the maximal eigenvalue of $J$. The unitary gate~\eqref{eq:R_matrix} is symmetric
\begin{align}
\label{eq:symmetric_R}
    [R^{s_1,s_2}(\lambda)]^T=R^{s_1,s_2}(\lambda),
\end{align}
and it satisfies the following normalization conditions:
\begin{align}
\label{eq:properties}
    R^{s_1,s_2}(\lambda)R^{s_1,s_2}(-\lambda)=\mathbbm{1},\quad
    \lim_{\lambda\to\infty}R^{s_1,s_2}(\lambda)=\mathbbm{1}.
\end{align}
In particular, the second property in Eq.~\eqref{eq:properties} implies that, for large $\lambda$, the unitary map $U=P^{s_1,s_2}R^{s_1,s_2}(\lambda)$ reduces to the permutation operator $P^{s_1,s_2}$.

We  will consider three instances of quantum ratchet circuits constructed from the $R$-matrix gate~\eqref{eq:R_matrix}:
\begin{enumerate}
    \item Setting $\lambda=\tau$ uniformly in all local gates $V=R^{s_1,s_2}(\tau)$, the dynamics is \emph{integrable}. Indeed, $V$ then corresponds precisely to the fused quantum $R$-matrix associated with an alternating spin chain~\cite{kulish1981,faddeev1996,bytsko2004}, whereas the full propagator~\eqref{eq:propagator} belongs to an infinite hierarchy of commuting transfer matrices (as detailed out in Section~\ref{sec:integrability} below).
    \item For a non-uniform choice of $\lambda$ staggered in time, i.e., with $\lambda=\pm\tau$ alternating between the adjacent horizontal layers of the lattice shown in Fig.~\ref{fig:ratchet_diagram}, the propagator~\eqref{eq:propagator} consists of two different, non-commuting transfer matrices. Hence, the resulting dynamics is not integrable.
    \item Regarding parameter $\tau$ of $R^{s_1,s_2}(\tau)$ as an independent and identically distributed random variable, we obtain a ``disordered'' (i.e., ``noisy'') and thus non-integrable circuit.
\end{enumerate} 

\subsection{$\mathcal{PT}$ symmetry}

The full quantum many-body spin ratchet depicted in Fig.~\ref{fig:ratchet_diagram} manifestly lacks symmetry under spatial reflection $\mathcal{P}$, $\ell\mapsto L-\ell+1$ (with the exception of the homogeneous lattice with equal spins $s_1=s_2$ which is not of our interest). Under the action of $\mathcal{P}$ the two spins get interchanged, $s_{1}\leftrightarrow s_{2}$, and consequently $\mathcal{P}(U)\neq U$ in the general case of $s_{1}\neq s_{2}$.  Similarly, the spin ratchet circuit breaks also the time-reversal symmetry $\mathcal{T}$. The latter is understood as an adjoint mapping $\mathcal{T}(\mathbb{U})\equiv SK\mathbb{U}K S^{-1}$ combining a unitary operator $S$ and an antiunitary conjugation $K$, such that
\begin{align}
    \mathcal{T}(\mathbb{U})=\mathbb{U}^{-1}.
\end{align}
We will show that the integrable version of the ratchet circuit with identical gates nonetheless obeys the joint $\mathcal{PT}$ symmetry. More generally, this is true for any brickwork ratchet circuit composed of identical gates~\eqref{eq:quantum_gate} in which the unitary $V$ is symmetric, e.g., see Eq.~\eqref{eq:symmetric_R}. The breaking of the space-reflection symmetry then naturally implies the breaking of the time-reversal symmetry.

To demonstrate that the integrable ratchet circuit is invariant under the $\mathcal{PT}$ symmetry, we first note that, for large $\tau$, the local unitary map $U$ becomes a permutation, $U(\tau\to \infty)=P^{s_{1} s_{2}}=\sum_{n_1=-s_1}^{s_1}\sum_{n_2=-s_2}^{s_2}\ket{n_2 n_1}\!\bra{n_1 n_2}$, for which the exchange of spins coincides with a matrix transposition, i.e.,
\begin{align}
\label{eq:parity_action}
    \mathcal{P}(U)=U^T.
\end{align} 
As it turns out, the same property still holds for any finite value of $\tau$ and,  in general, for any symmetric unitary gate $V$ in Eq.~\eqref{eq:quantum_gate}---see Appendix~\ref{app:pt_invariance}. To find the inverse $\mathbb{U}^{-1}=\mathbb{U}^{-1}_{\rm o}\mathbb{U}^{-1}_{\rm e}$ of the full propagator $\mathbb{U}=\mathbb{U}_{\rm e}\mathbb{U}_{\rm o}$, one can thus make use of the following sequence of  transformations: 
\begin{enumerate}
    \item 
    spatial reflection $\mathcal{P}$ (acting as a transposition---see Eq.~\ref{eq:parity_action}) at the level of local unitary gates;
    \item a conjugation with an antiunitary matrix $K$, $KUK=U^*$, which, in combination with $\mathcal{P}$, inverts the unitary gates, namely $K\mathcal{P}(U)K=U^{-1}$;
    \item a one-site lattice shift, exchanging the order of the two consecutive time steps: $\mathbb{U}_{\rm e}^{-1}\mathbb{U}_{\rm o}^{-1}\mapsto \mathbb{U}_{\rm o}^{-1}\mathbb{U}_{\rm e}^{-1}$.
\end{enumerate}
The composition of the last two transformations constitutes an antiunitary map which may be regarded as the time-reversal transformation $\mathcal{T}$. In summary, ratchet circuits composed of identical unitary gates~\eqref{eq:quantum_gate} with $V^T=V$, a particular example being the integrable one, represent $\mathcal{PT}$-symmetric systems that lack both $\mathcal{P}$ and $\mathcal{T}$ symmetries (see Appendix~\ref{app:pt_invariance} for a detailed proof of the $\mathcal{PT}$ symmetry).

\section{Integrable quantum ratchet}
\label{sec:integrability}

In this section we detail out the structure of integrable quantum ratchets. To this end, we fix all free unitary gate parameters $\lambda$ to the same value $\tau\in\mathbb{R}$. Integrability then follows from the fact that the $R$-matrix~\eqref{eq:R_matrix} satisfies the Yang-Baxter equation, 
\begin{align}
\label{eq:yang-baxter}
    R_{1,2}^{s_1,s_2}&(\lambda-\mu)R_{1,3}^{s_1,s_3}(\lambda)R_{2,3}^{s_2,s_3}(\mu)=\notag\\
    &=R_{2,3}^{s_2,s_3}(\mu)R_{1,3}^{s_1,s_3}(\lambda)R_{1,2}^{s_1,s_2}(\lambda-\mu),
\end{align}
over a three-fold product space $\mathcal{H}_{s_1}\otimes\mathcal{H}_{s_2}\otimes\mathcal{H}_{s_3}$, for an arbitrary triple of integer or half-integer spins $\{s_k\}_{k=1}^3$, and for any two complex parameters $\lambda,\mu\in\mathbb{C}$~\cite{kulish1981,faddeev1996,bytsko2004}. As shown in Appendix~\ref{app:integrability}, the Yang-Baxter equation leads to a family of commuting transfer matrices which includes the full time-step propagator~\eqref{eq:propagator} as a particular instance. Moreover, such transfer matrices can be simultaneously diagonalized using the algebraic Bethe ansatz~\cite{faddeev1996}. Leaving the technical details of this procedure to Appendix~\ref{app:integrability}, we here outline a simple way of establishing the existence of commuting transfer matrices for an integrable ratchet circuit depicted in Fig.~\ref{fig:ratchet_diagram}. As our starting point, we consider the Yang-Baxter equation~\eqref{eq:yang-baxter} multiplied from the left-hand side by $P^{s_1,s_2}_{1,2}$. Then, by introducing $\lambda_{\pm}=\lambda\pm\tau/2$ and recognizing the quantum gate~\eqref{eq:quantum_gate} acting on the sites $1$ and $2$, we have the identity
\begin{align}
\label{eq:permuted_yang-baxter}
    U^{}_{1,2}R^{s_1\!,s_3}_{1,3}\!(\lambda_+\!)R^{s_2\!,s_3}_{2,3}\!(\lambda_-\!)\!=\!R^{s_2\!,s_3}_{1,3}\!(\lambda_-\!)R^{s_1\!,s_3}_{2,3}\!(\lambda_+\!)U^{}_{1,2},
\end{align}
or pictorially, in terms of diagrams,
\begin{align}
\label{eq:yang-baxter_diagram}
\centering
\begin{tikzpicture}[baseline=(current  bounding  box.center),scale=1]
     \draw[black,opacity=0.075,line width=0.2mm] (-2.25+0.075,-0.4-0.075) -- (-2.25+0.075,-0.075-0.075);
     \draw[black,opacity=0.075,line width=0.6mm] (-1.25+0.075,-0.4-0.075) -- (-1.25+0.075,-0.075-0.075);
     \draw[black,opacity=0.075,line width=0.2mm] (-1.975+0.075,1.025-0.075) to [in=90,out=225] (-2.25+0.075,0.575-0.075);
     \draw[black,opacity=0.075,line width=0.6mm] (-1.525+0.075,1.025-0.075) to [in=90,out=-45] (-1.25+0.075,0.575-0.075);
     \draw[black,opacity=0.075,line width=0.2mm] (-1.525+0.075,1.475-0.075) -- (-1.25+0.075,1.75-0.075);
     \draw[black,opacity=0.075,line width=0.6mm] (-1.975+0.075,1.475-0.075) -- (-2.25+0.075,1.75-0.075);
     \draw[black,opacity=0.075,line width=0.4mm] (-2.925+0.075,0.25-0.075) -- (-2.575+0.075,0.25-0.075);
     \draw[black,opacity=0.075,line width=0.4mm] (-1.925+0.075,0.25-0.075) -- (-1.575+0.075,0.25-0.075);
     \draw[black,opacity=0.075,line width=0.4mm] (-0.925+0.075,0.25-0.075) -- (-0.575+0.075,0.25-0.075);
     \draw[black,opacity=0.075,line width=0.6mm] (1.75-0.5+0.075,1.575-0.075) -- (1.75-0.5+0.075,1.9-0.075);
     \draw[black,opacity=0.075,line width=0.2mm] (2.75-0.5+0.075,0.925-0.075) to [in=45,out=-90] (2.475-0.5+0.075,0.475-0.075);
     \draw[black,opacity=0.075,line width=0.6mm] (1.75-0.5+0.075,0.925-0.075) to [in=135,out=-90] (2.025-0.5+0.075,0.475-0.075);
     \draw[black,opacity=0.075,line width=0.2mm] (2.75-0.5+0.075,1.575-0.075) -- (2.75-0.5+0.075,1.9-0.075);
     \draw[black,opacity=0.075,line width=0.6mm] (2.475-0.5+0.075,0.025-0.075) -- (2.75-0.5+0.075,-0.25-0.075);
     \draw[black,opacity=0.075,line width=0.2mm] (2.025-0.5+0.075,0.025-0.075) -- (1.75-0.5+0.075,-0.25-0.075);
     \draw[black,opacity=0.075,line width=0.4mm] (-2.925+4-0.5+0.075,1.25-0.075) -- (-2.5 75+4-0.5+0.075,1.25-0.075);
     \draw[black,opacity=0.075,line width=0.4mm] (-1.925+4-0.5+0.075,1.25-0.075) -- (-1.575+4-0.5+0.075,1.25-0.075);
     \draw[black,opacity=0.075,line width=0.4mm] (-0.925+4-0.5+0.075,1.25-0.075) -- (-0.575+4-0.5+0.075,1.25-0.075);
     \fill[white!90!black, opacity=1, rounded corners = 2] (-2+0.075,1-0.075) rectangle ++(0.5,0.5);
     \fill[rotate around={45:(-2.25+0.075,0.25-0.075))},white!90!black, opacity=1, rounded corners = 2] (-2.5+0.075,0-0.075) rectangle ++(0.5,0.5);
     \fill[rotate around={45:(-1.25+0.075,0.25-0.075))},white!90!black, opacity=1, rounded corners = 2] (-1.5+0.075,0-0.075) rectangle ++(0.5,0.5);
     \fill[white!90!black, opacity=1, rounded corners = 2] (2-0.5+0.075,0-0.075) rectangle ++(0.5,0.5);
     \fill[rotate around={45:(1.75-0.5+0.075,1.25-0.075))}, white!90!black, opacity=1, rounded corners = 2,thick] (1.5-0.5+0.075,1-0.075) rectangle ++(0.5,0.5);
     \fill[rotate around={45:(2.75-0.5+0.075,1.25-0.075))}, white!90!black, opacity=1, rounded corners = 2,thick] (2.5-0.5+0.075,1-0.075) rectangle ++(0.5,0.5);
     \draw[fill=red!10!green!70!blue!20, opacity=1, rounded corners = 2,thick] (-2,1) rectangle ++(0.5,0.5);
     \node[anchor=center] at (-1.75,1.25) {$\tau$};
     \draw[rotate around={45:(-2.25,0.25))},fill=blue!20!white, opacity=1, rounded corners = 2,thick] (-2.5,0) rectangle ++(0.5,0.5);
     \node[anchor=center] at (-2.25,0.25) {$\lambda_+$};
     \draw[rotate around={45:(-1.25,0.25))},fill=red!20!white, opacity=1, rounded corners = 2,thick] (-1.5,0) rectangle ++(0.5,0.5);
     \node[anchor=center] at (-1.25,0.25) {$\lambda_-$};
     \draw[fill=red!10!green!70!blue!20, opacity=1, rounded corners = 2,thick] (2-0.5,0) rectangle ++(0.5,0.5);
     \node[anchor=center] at (2.25-0.5,0.25) {$\tau$};
     \draw[rotate around={45:(1.75-0.5,1.25))},fill=red!20!white, opacity=1, rounded corners = 2,thick] (1.5-0.5,1) rectangle ++(0.5,0.5);
     \node[anchor=center] at (1.75-0.5,1.25) {$\lambda_-$};
     \draw[rotate around={45:(2.75-0.5,1.25))},fill=blue!20!white, opacity=1, rounded corners = 2,thick] (2.5-0.5,1) rectangle ++(0.5,0.5);
     \node[anchor=center] at (2.75-0.5,1.25) {$\lambda_+$};
     \node[anchor=center] at (0,0.75) {$=\phantom{,}$};
     \node[anchor=center] at (3.375,0.75) {$\phantom{=},$};
     \draw[black,line width=0.2mm] (-2.25,-0.4) -- (-2.25,-0.075);
     \draw[blue!60!white,line width=0.6mm] (-1.25,-0.4) -- (-1.25,-0.075);
     \draw[black,line width=0.2mm] (-1.975,1.025) to [in=90,out=225] (-2.25,0.575);
     \draw[blue!60!white,line width=0.6mm] (-1.525,1.025) to [in=90,out=-45] (-1.25,0.575);
     \draw[black,line width=0.2mm] (-1.525,1.475) -- (-1.25,1.75);
     \draw[blue!60!white,line width=0.6mm] (-1.975,1.475) -- (-2.25,1.75);
     \draw[red!60!white,line width=0.4mm] (-2.925,0.25) -- (-2.575,0.25);
     \draw[red!60!white,line width=0.4mm] (-1.925,0.25) -- (-1.575,0.25);
     \draw[red!60!white,line width=0.4mm] (-0.925,0.25) -- (-0.575,0.25);
     \draw[blue!60!white,line width=0.6mm] (1.75-0.5,1.575) -- (1.75-0.5,1.9);
     \draw[black,line width=0.2mm] (2.75-0.5,0.925) to [in=45,out=-90] (2.475-0.5,0.475);
     \draw[blue!60!white,line width=0.6mm] (1.75-0.5,0.925) to [in=135,out=-90] (2.025-0.5,0.475);
     \draw[black,line width=0.2mm] (2.75-0.5,1.575) -- (2.75-0.5,1.9);
     \draw[blue!60!white,line width=0.6mm] (2.475-0.5,0.025) -- (2.75-0.5,-0.25);
     \draw[black,line width=0.2mm] (2.025-0.5,0.025) -- (1.75-0.5,-0.25);
     \draw[red!60!white,line width=0.4mm] (-2.925+4-0.5,1.25) -- (-2.5 75+4-0.5,1.25);
     \draw[red!60!white,line width=0.4mm] (-1.925+4-0.5,1.25) -- (-1.575+4-0.5,1.25);
     \draw[red!60!white,line width=0.4mm] (-0.925+4-0.5,1.25) -- (-0.575+4-0.5,1.25);
     \node[anchor=north] at (-2.25,-0.4) {$s_1$};
     \node[anchor=north] at (-1.25,-0.4) {$s_2$};
     \node[anchor=east] at (-2.925,0.25) {$s_3$};
     \node[anchor=south] at (-2.25+4-0.5,1.9) {$s_2$};
     \node[anchor=south] at (-1.25+4-0.5,1.9) {$s_1$};
     \node[anchor=west] at (3.425-0.5,1.25) {$s_3$};
\end{tikzpicture}
\end{align}
where $\tau$ is a free parameter of the unitary map $U$ and $\lambda_\pm$ are the spectral parameters of the two $R$-matrices. Using this diagram twice, applying it from below in Fig.~\ref{fig:ratchet_diagram}, one can straightforwardly verify that the entire ratchet circuit commutes with a row transfer matrix of the form
\begin{align}
\label{eq:transfer_matrix_diagram}
    \centering
    \begin{tikzpicture}[baseline=(current  bounding  box.center),scale=1]
    \draw[black,opacity=0.075,line width=0.4mm] (-2.75+0.075,0-0.075) to[out=180,in=-90] (-2.95+0.075,0.2-0.075);
    \draw[black,opacity=0.075,line width=0.4mm]  (-2.95+0.075,0.2-0.075) to[out=90,in=180] (-2.75+0.075,0.4-0.075);
    \draw[black,opacity=0.075,line width=0.4mm] (-2.75+0.075,0-0.075) -- (-2.6+0.075,0-0.075);
    \draw[black,opacity=0.075,line width=0.4mm] (-1.95+0.075,0-0.075) -- (-1.6+0.075,0-0.075);
    \draw[black,opacity=0.075,line width=0.4mm] (-0.95+0.075,0-0.075) -- (-0.6+0.075,0-0.075);
    \draw[black,opacity=0.075,line width=0.4mm] (0.05+0.075,0-0.075) -- (0.2+0.075,0-0.075);
    \draw[black,opacity=0.075,line width=0.4mm] (1.25+0.075,0-0.075) -- (1.4+0.075,0-0.075);
    \draw[black,opacity=0.075,line width=0.4mm] (2.05+0.075,0-0.075) -- (2.4+0.075,0-0.075);
    \draw[black,opacity=0.075,line width=0.4mm] (3.05+0.075,0-0.075) -- (3.2+0.075,0-0.075);
    \draw[black,opacity=0.075,line width=0.4mm] (3.2+0.075,0-0.075) to[out=0,in=-90] (3.4+0.075,0.2-0.075);
    \draw[black,opacity=0.075,line width=0.4mm] (3.4+0.075,0.2-0.075) to[out=90,in=0] (3.2+0.075,0.4-0.075);
    \draw[black,opacity=0.075,line width=0.2mm] (-2.275+0.075,-0.325-0.075) -- (-2.275+0.075,-0.625-0.075);
    \draw[black,opacity=0.075,line width=0.6mm] (-1.275+0.075,-0.325-0.075) -- (-1.275+0.075,-0.625-0.075);
    \draw[black,opacity=0.075,line width=0.2mm] (-0.275+0.075,-0.325-0.075) -- (-0.275+0.075,-0.625-0.075);
    \draw[black,opacity=0.075,line width=0.2mm] (-2.275+0.075,0.325-0.075) -- (-2.275+0.075,0.625-0.075);
    \draw[black,opacity=0.075,line width=0.6mm] (-1.275+0.075,0.325-0.075) -- (-1.275+0.075,0.625-0.075);
    \draw[black,opacity=0.075,line width=0.2mm] (-0.275+0.075,0.325-0.075) -- (-0.275+0.075,0.625-0.075);
    \draw[black,opacity=0.075,line width=0.6mm] (2.725+0.075,-0.325-0.075) -- (2.725+0.075,-0.625-0.075);
    \draw[black,opacity=0.075,line width=0.2mm] (1.725+0.075,-0.325-0.075) -- (1.725+0.075,-0.625-0.075);
    \draw[black,opacity=0.075,line width=0.6mm] (2.725+0.075,0.325-0.075) -- (2.725+0.075,0.625-0.075);
    \draw[black,opacity=0.075,line width=0.2mm] (1.725+0.075,0.325-0.075) -- (1.725+0.075,0.625-0.075);
    \fill[rotate around={45:(-2.275+0.075,0-0.075))},white!90!black, opacity=1, rounded corners = 2] (-2.525+0.075,-0.25-0.075) rectangle ++(0.5,0.5);
    \fill[rotate around={45:(-1.275+0.075,0-0.075))},white!90!black, opacity=1, rounded corners = 2] (-1.525+0.075,-0.25-0.075) rectangle ++(0.5,0.5);
    \fill[rotate around={45:(-0.275+0.075,0-0.075))},white!90!black, opacity=1, rounded corners = 2] (-0.525+0.075,-0.25-0.075) rectangle ++(0.5,0.5);
    \fill[rotate around={45:(-2.275+4+0.075,0-0.075))},white!90!black, opacity=1, rounded corners = 2] (-2.525+4+0.075,-0.25-0.075) rectangle ++(0.5,0.5);
    \fill[rotate around={45:(-1.275+4+0.075,0-0.075))},white!90!black, opacity=1, rounded corners = 2] (-1.525+4+0.075,-0.25-0.075) rectangle ++(0.5,0.5);
    \draw[red!60!white,line width=0.4mm] (-2.75,0) to[out=180,in=-90] (-2.95,0.2);
    \draw[red!60!white,line width=0.4mm]  (-2.95,0.2) to[out=90,in=180] (-2.75,0.4);
    \draw[red!60!white,line width=0.4mm] (-2.75,0) -- (-2.6,0);
    \draw[red!60!white,line width=0.4mm] (-1.95,0) -- (-1.6,0);
    \draw[red!60!white,line width=0.4mm] (-0.95,0) -- (-0.6,0);
    \draw[red!60!white,line width=0.4mm] (0.05,0) -- (0.2,0);
    \draw[red!60!white,line width=0.4mm] (1.25,0) -- (1.4,0);
    \draw[red!60!white,line width=0.4mm] (2.05,0) -- (2.4,0);
    \draw[red!60!white,line width=0.4mm] (3.05,0) -- (3.2,0);
    \draw[red!60!white,line width=0.4mm] (3.2,0) to[out=0,in=-90] (3.4,0.2);
    \draw[red!60!white,line width=0.4mm] (3.4,0.2) to[out=90,in=0] (3.2,0.4);
    \node[anchor=center] at (0.725,0) {$\ldots$};
    \draw[rotate around={45:(-2.275,0))},fill=blue!20!white, opacity=1, rounded corners = 2,thick] (-2.525,-0.25) rectangle ++(0.5,0.5);
    \draw[rotate around={45:(-1.275,0))},fill=red!20!white, opacity=1, rounded corners = 2,thick] (-1.525,-0.25) rectangle ++(0.5,0.5);
    \draw[rotate around={45:(-0.275,0))},fill=blue!20!white, opacity=1, rounded corners = 2,thick] (-0.525,-0.25) rectangle ++(0.5,0.5);
    \draw[rotate around={45:(-2.275+4,0))},fill=blue!20!white, opacity=1, rounded corners = 2,thick] (-2.525+4,-0.25) rectangle ++(0.5,0.5);
    \draw[rotate around={45:(-1.275+4,0))},fill=red!20!white, opacity=1, rounded corners = 2,thick] (-1.525+4,-0.25) rectangle ++(0.5,0.5);
    \node[anchor=center] at (-2.25,0) {$\lambda_+$};
    \node[anchor=center] at (-1.25,0) {$\lambda_-$};
    \node[anchor=center] at (-0.25,0) {$\lambda_+$};
    \node[anchor=center] at (-0.25+2,0) {$\lambda_+$};
    \node[anchor=center] at (0.75+2,0) {$\lambda_-$};
    \node[anchor=center] at (3.5,0) {$\phantom{=},$};
    \draw[black,line width=0.2mm] (-2.275,-0.325) -- (-2.275,-0.625);
    \draw[blue!60!white,line width=0.6mm] (-1.275,-0.325) -- (-1.275,-0.625);
    \draw[black,line width=0.2mm] (-0.275,-0.325) -- (-0.275,-0.625);
    \draw[black,line width=0.2mm] (-2.275,0.325) -- (-2.275,0.625);
    \draw[blue!60!white,line width=0.6mm] (-1.275,0.325) -- (-1.275,0.625);
    \draw[black,line width=0.2mm] (-0.275,0.325) -- (-0.275,0.625);
    \draw[blue!60!white,line width=0.6mm] (2.725,-0.325) -- (2.725,-0.625);
    \draw[black,line width=0.2mm] (1.725,-0.325) -- (1.725,-0.625);
    \draw[blue!60!white,line width=0.6mm] (2.725,0.325) -- (2.725,0.625);
    \draw[black,line width=0.2mm] (1.725,0.325) -- (1.725,0.625);
    \node[anchor=north] at (-2.275,-0.625) {$s_1$};
    \node[anchor=north] at (-1.275,-0.625) {$s_2$};
    \node[anchor=north] at (-0.275,-0.625) {$s_1$};
    \node[anchor=north] at (1.725,-0.625) {$s_1$};
    \node[anchor=north] at (2.725,-0.625) {$s_2$};
    \end{tikzpicture}
\end{align}
where the horizontal red line encloses a loop, indicating the partial trace over an (auxiliary) space $\mathcal{H}_{s_{3}}$ of spin $s_3$. We have thus established commutativity,
\begin{align}
    [\,\mathbb{U},T_{s_3}(\lambda)\,]=0,    
\end{align}
of the propagator~\eqref{eq:propagator} with a staggered transfer matrix
\begin{align}
\label{eq:transfer_matrix}
    T_{s_3}(\lambda)={\rm Tr}_{a}\prod_{1\le j\le L/2}^{\rightarrow}R^{s_1,s_3}_{2j-1,a}(\lambda_+)R^{s_2,s_3}_{2j,a}(\lambda_-),
\end{align}
graphically represented in the diagram~\eqref{eq:transfer_matrix_diagram}. The arrow direction on top of the product specifies the spatial ordering of the matrix product. In our convention, the physical (lower) indices of the $R$-matrices increase from the left-hand towards the right-hand side. 

\subsection{Magnon dispersion relation}

A distinguished role is played by the transfer matrices in which the auxiliary space corresponds to either of the two alternating spins in the chain, that is $s_3 \in \{s_1,s_2\}$. In particular, in Appendix~\ref{app:integrability} we show that 
\begin{equation}
\label{eq:odd_even_part}
    T_{s_2}(\tfrac{\tau}{2})=\mathbb{U}^{}_{\rm e}\Pi_{s_1, s_2},\quad
    T_{s_1}(-\tfrac{\tau}{2})=\mathbb{U}^{-1}_{\rm o} \Pi_{s_1, s_2},
\end{equation}
where we have defined a one-site lattice shift in the backward direction, $\Pi_{s_1,s_2}\!:\!(\mathcal{H}_{s_1}\!\otimes\!\mathcal{H}_{s_2})^{\otimes L/2}\!\to\!(\mathcal{H}_{s_2}\!\otimes\!\mathcal{H}_{s_1})^{\otimes L/2}$, reading
\begin{equation}
\label{eq:shift}
    \Pi_{s_1,s_2}=P^{s_1,s_2}_{1,2}P^{s_2,s_1}_{1,3}P^{s_1,s_2}_{1,4}P^{s_2,s_1}_{1,5}\cdots P^{s_1,s_2}_{1,L}.
\end{equation} 
In analogy with the light-cone lattice discretizations of certain integrable quantum field theories~\cite{destri1987,destri1989,volkov1992,faddeev1996}, Eqs.~\eqref{eq:odd_even_part} can be interpreted as elementary lattice shift operators along the light-cone directions, i.e., the north-west and south-west direction, respectively. In this view, the propagator $\mathbb{U}$ realizes a two-site lattice shift in the ``time'' (i.e., north) direction, and similarly we introduce $\mathbb{T}$ as a two-site lattice shift in the backward (i.e., west) spatial direction:
\begin{align}
\begin{aligned}
\label{eq:propagator_from_transfer_matrices}
    \mathbb{U}&=T_{s_2}(\tfrac{\tau}{2})\left[T_{s_1}(-\tfrac{\tau}{2})\right]^{-1},\\
    \mathbb{T}&=T_{s_2}(\tfrac{\tau}{2})T_{s_1}(-\tfrac{\tau}{2}).
\end{aligned}
\end{align}
Expressing the lattice shifts in terms of transfer matrices allows us to infer their eigenvalues using the algebraic Bethe ansatz. In particular, the unimodular eigenvalues of $\mathbb{T}$ and $\mathbb{U}$ correspond to quasimomentum and quasienergy, respectively.

Eigenstates of integrable quantum spin chains can be described in terms of elementary spin-wave excitations called magnons. An eigenstate involving $N$ magnons is parametrized by a set of rapidities, $\{\lambda_j\}_{j=1}^{N}$. To ensure periodicity of the wavefunction, the rapidities have to obey the Bethe equations, which impose the condition that the total phase acquired by a quasiparticle upon scattering with other quasiparticles, while traversing the spin chain, is trivial. They read
\begin{align}
\label{eq:bethe_equations}
    e^{i L p(\lambda_j)}\prod^N_{\substack{k=1\\k\ne j}}\mathcal{S}(\lambda_j-\lambda_k)=1,
\end{align}
where $\mathcal{S}(\lambda-\mu)=(\lambda-\mu+i)/(\lambda-\mu-i)$ denotes the scattering amplitude associated with a two-magnon scattering and $p(\lambda)$ is the single-magnon quasimomentum. In terms of the single-magnon quasimomentum $p^{(2s)}$ of a homogeneous Heisenberg spin-$s$ chain,
\begin{align}
\label{eq:single-particle-momentum}
    p^{(2s)}(\lambda)\equiv i\log\left(\frac{\lambda\!+\!i s}{\lambda\!-\!i s}\right),
\end{align}
which (for notational convenience) we label by an integer number $2s\in\mathbb{N}$, the single-magnon quasimomentum $p(\lambda)$ in a ratchet circuit decomposes as
\begin{align}
    \label{eq:quasimomentum} 
    p(\lambda)=\frac{1}{2}\left[p^{(2s_1)}(\lambda_+)+p^{(2s_2)}(\lambda_-)\right].
\end{align}
This implies that the two-site lattice shift in the backward spatial direction acts on the eigenstates as 
\begin{align}
\label{eq:T-eigenvalue}
    \mathbb{T}\ket{\{\lambda_j\}}&=e^{-2i \sum_{j=1}^N p(\lambda_j)}\ket{\{\lambda_j\}},
\end{align}
where we have singled out a factor of $2$ in the eigenvalue, in order to associate the total quasimomentum $\sum_{j=1}^N p(\lambda_j)$ with a one-site lattice shift. For the propagator in the temporal direction we obtain a similar expression
\begin{align}
\label{eq:U-eigenvalue}
    \mathbb{U}\ket{\{\lambda_j\}}&=e^{i 2\sum_{j=1}^N \varepsilon(\lambda_j)}\ket{\{\lambda_j\}},
\end{align}
where $\varepsilon(\lambda)$ denotes the single-magnon quasienergy (see Appendix~\ref{app:integrability}). Analogously to the single-magnon quasimomentum~\eqref{eq:quasimomentum}, we find $\varepsilon(\lambda)$ to be the difference of the single-magnon quasimomenta of the homogeneous chains with different spins, namely
\begin{align}
    \label{eq:quasienergy}
    \varepsilon(\lambda)=\frac{1}{2}\left[p^{(2s_1)}(\lambda_+)-p^{(2s_2)}(\lambda_-)\right].
\end{align} 

\subsection{Homogeneous chain and integrable Trotterization}

In the case of a homogeneous chain with $s_1=s_2=s$, the $R$-matrix obeys an additional property $R^{s,s}(0)=P^{s,s}$. For small values of $\tau$, we can then expand the quantum gate $U=P^{s,s}R^{s,s}(\tau)$ around the identity. In the scaling limit $\tau\to 0$ and $t\to\infty$, with the product $t\tau$ fixed, the Floquet dynamics with the propagator~\eqref{eq:propagator} therefore yields a continuous time-evolution with a time parameter proportional to $t\tau$. Such a configuration of quantum gates is a quantum circuit corresponding to an integrable Trotterization of the continuous-time dynamics generated by the Heisenberg spin-$s$ model~\cite{vanicat2018,ljubotina2019}. This highlights that our approach can be seen as a nontrivial generalization of unitary circuits used in state-of-the-art quantum computation experiments~\cite{morvan2022,maruyoshi2023,rosenberg2024}.

Expanding the single-magnon quasienergy~\eqref{eq:quasienergy} in $\tau$ up to the leading order, we find
\begin{align}
\label{eq:continuous-time_energy}
    \varepsilon(\lambda)=\frac{s}{\lambda^2+s^2}\tau+O(\tau^2).
\end{align}
Similarly, the single-magnon quasimomentum~\eqref{eq:quasimomentum} becomes simply $\lim_{\tau\to 0}p(\lambda)=p^{(2s)}(\lambda)$.
In the Trotter limit we thus recover the standard relation $\varepsilon^{(2s)}(\lambda)=\tfrac{1}{2}\partial_\lambda p^{(2s)}(\lambda)$, where we have defined $\varepsilon^{(2s)}(\lambda)\equiv\lim_{\tau\to 0}[\varepsilon(\lambda)/\tau]$ (see, e.g., Ref.~\cite{faddeev1996}).

\subsection{Semiclassical limit}

The integrable quantum many-body spin ratchet admits a semiclassical limit. Instead of local unitary gates $U$, the classical version of the many-body spin ratchet is made out of SO(3)-symmetric symplectic two-body maps $\Phi_{\tau}$ that depend on a time-step parameter $\tau \in \mathbb{R}_{+}$. To ensure integrability, we introduce a classical Lax operator ${\rm L}(\boldsymbol{S}, \lambda)$
as a matrix-valued function on the local phase space $\mathcal{S}_{r}$ (a $2$-sphere of radius $r$) of the \emph{classical} spin $\boldsymbol{S}$ of length $r$.
In terms of Pauli matrices
$\boldsymbol{\sigma} = (\sigma^x, \sigma^y, \sigma^z)$, the Lax matrix takes the form
\begin{equation}
{\rm L}(\boldsymbol{S}, \lambda) = \frac{2i\lambda \mathbbm{1}  + \boldsymbol{S} \cdot \boldsymbol{\sigma}}{2i\lambda  + r}.
\end{equation}
Upon replacing the $R$-matrices in Eq.~\eqref{eq:permuted_yang-baxter} with classical Lax operators, setting $s_{1, 2} = r_{1, 2}s$, and subsequently taking the $s \to \infty$ limit, conjugation with the gate $U$ becomes equivalent to the symplectic map $\Phi_\tau: \mathcal{S}_{r_1} \times \mathcal{S}_{r_2} \to \mathcal{S}_{r_2} \times \mathcal{S}_{r_1}$. The latter acts as $(\boldsymbol{S}'_1, \boldsymbol{S}'_2) = \Phi_\tau(\boldsymbol{S}_1, \boldsymbol{S}_2)$, where
\begin{align}
\begin{aligned}
\label{classical_map}
    \boldsymbol{S}'_{1} &= \frac{(\sigma^2 - \eta^2) \boldsymbol{S}_{1} + (\tau^2 - \eta^2) \boldsymbol{S}_{2} + \tau \boldsymbol{S}_{1}\times \boldsymbol{S}_{2}}{\tau^2 + \sigma^2},\\
    \boldsymbol{S}'_{2} &= \frac{(\sigma^2 - \eta^2) \boldsymbol{S}_{2} + (\tau^2 - \eta^2) \boldsymbol{S}_{1} + \tau \boldsymbol{S}_{2}\times \boldsymbol{S}_{1}}{\tau^2 + \sigma^2},
\end{aligned}
\end{align}
with $\sigma^2 = (\boldsymbol{S}_1 + \boldsymbol{S}_2)^2/4$ and $\eta^2 = (\boldsymbol{S}_1 + \boldsymbol{S}_2)(\boldsymbol{S}_1 - \boldsymbol{S}_2)/4 = (r_1^2 - r_2^2)/4$. The Yang-Baxter equation~\eqref{eq:permuted_yang-baxter} thus becomes equivalent to the discrete zero-curvature condition
\begin{align}
    {\rm L}(\boldsymbol{S}_1, \lambda_+){\rm L}(\boldsymbol{S}_2, \lambda_-) = {\rm L}(\boldsymbol{S}'_1, \lambda_-){\rm L}(\boldsymbol{S}'_2, \lambda_+).
\end{align}

For spins of equal length $r_{1,2}=1$, the semiclassical limit was derived in Ref.~\cite{sklyanin1988}, and the resulting dynamics has been investigated in Ref.~\cite{Krajnik2020} (see also Refs.~\cite{Krajnik2020a,krajnik2021}).

\vspace{2em}
\section{Transport properties and hydrodynamics}
\label{sec:hydro}

Chiral spin dynamics at the microscopic level profoundly affects transport properties on a macroscopic scale. To set the ground for their investigation, we first introduce the conserved U(1) charge (i.e., total magnetization) and the associated current. We then proceed by exploring the consequences of broken space-reflection symmetry using the tools of the thermodynamic Bethe ansatz. By employing generalized hydrodynamics~\cite{alvaredo2016,bertini2016}, we examine the properties of the dynamical two-point correlation function of the magnetization density. Afterwards we investigate the structure of large-scale current fluctuations~\cite{bernard2013,doyon2023} in the stationary maximum-entropy ensembles and discuss the impact of the time-reversal symmetry breaking on the associated large-deviation rate function.

\subsection{Magnetization density and current} 
\label{sec:continuity_equations}

The magnitude of the total spin $J$, entering the
unitary gate~\eqref{eq:quantum_gate} through the $R$-matrix~\eqref{eq:R_matrix}, commutes with the total projection of the two-site magnetization $S^z_1+S^z_2$ onto the $z$-axis. Accounting for the additional permutation, we thus have the relation
\begin{align}
\label{eq:permuted_U1}
    U_{\ell\!-\!1,\ell}^{-\!1}\left[S^z_{2;\ell\!-\!1}\!+\!S^z_{1;\ell}\right]U^{}_{\ell\!-\!1,\ell}=S^z_{1;\ell\!-\!1}\!+\!S^z_{2;\ell},
\end{align}
where the extra lower indices $\ell-1$ and $\ell$ designate the lattice sites on which the local spin densities act nontrivially. For equal spins, $s_1=s_2$, Eq.~\eqref{eq:permuted_U1} implies conservation of local magnetization along the $z$-axis at the level of individual quantum gates. In the general case with $s_1 \neq s_2$, however, only the total magnetization 
\begin{align}
\label{eq:magnetization}
    Q=\sum_{\ell=1}^{L/2}\left(S^z_{1;2\ell\!-\!1}\!+\!S^z_{2;2\ell}\right)
\end{align}
remains globally conserved, representing a global U(1) conserved charge of the ratchet circuit. Owing to the staggered structure of the circuit, there are in fact two independent local continuity equations associated with it~\cite{ljubotina2019}. In particular, defining magnetization densities on odd and even bonds as 
\begin{align}
\begin{aligned}
\label{eq:local_mag}
    q^{(\rm o)}_{2\ell-1}&=S^z_{1;2\ell-1}+S^z_{2;2\ell},\\
    q^{(\rm e)}_{2\ell}&=S^z_{2;2\ell}+S^z_{1;2\ell+1},
\end{aligned}
\end{align}
respectively, the associated magnetization current densities $j^{(\rm o)}$ and $j^{(\rm e)}$ satisfy the following continuity equations
\begin{align}
\begin{aligned}
\label{eq:contiuity_equations}
    \mathbb{U}^{-1}q^{(\rm e)}_{2\ell}\mathbb{U}-q^{(\rm e)}_{2\ell}&=-j^{(\rm e)}_{2\ell+1}+j^{(\rm e)}_{2\ell-1},\\
    \mathbb{U}^{-1}q^{(\rm o)}_{2\ell-1}\mathbb{U}-q^{(\rm o)}_{2\ell-1}&=-j^{(\rm o)}_{2\ell}+j^{(\rm o)}_{2\ell-2}.
\end{aligned}
\end{align}
Using Eq.~\eqref{eq:permuted_U1}, the first continuity equation~\eqref{eq:contiuity_equations} implies
\begin{align}
    j^{(\rm e)}_{2\ell-1}&=U^{-1}_{2\ell-1,2\ell}S^z_{1;2\ell}U^{}_{2\ell-1,2\ell}-S^z_{2;2\ell}\notag\\
    &=S^z_{1;2\ell-1}-U^{-1}_{2\ell-1,2\ell}S^z_{2;2\ell-1}U^{}_{2\ell-1,2\ell},
\end{align}
while the second one yields $j^{(\rm o)}_{2\ell}=\mathbb{U}_{\rm o}^{-1}j^{(\rm e)}_{2\ell}\mathbb{U}_{\rm o}^{}$. Explicit expressions for the spin current densities are rather cumbersome and we do not report them here.

\subsection{Dynamical structure factor and drift velocity}
\label{sec:drift_velocity}

To characterize spin transport, we investigate the hydrodynamic relaxation of the dynamical spin susceptibility (structure factor)~\cite{denardis2022}
\begin{align}
\label{eq:structure-factor}
    S(x,t)\equiv\langle q(x,t)q(0,0)\rangle^c,
\end{align}
where $\langle\bullet\rangle^c$ denotes the connected part of the correlation function and $q(x,t)$ denotes the time-evolved density of the total conserved magnetization $Q=\int{\rm d}x \, q(x)$. For convenience, we have passed from lattice to continuous space-time variables (omitting the precise identification for the time being). The local continuity equation therefore reads 
\begin{align}
\label{eq:charge-current}
    \partial_t q (x,t)+\partial_x j (x,t)=0,
\end{align}
where $j(x,t)$ is the current density associated with $q(x,t)$. By assuming that the ensemble averages of $q(x,t)$ vary slowly on large spatio-temporal scales, the late-time relaxation of $S(x,t)$ can be computed with the aid of hydrodynamics. In what follows, we will consider Gibbs states at a finite magnetization density set by the chemical potential $\mu$. Note that, since ratchet circuits are Floquet-driven systems in which temperature is not defined, such states are of the form 
\begin{align}
\label{eq:Gibbs-state}
    \varrho_\mu\equiv \frac{e^{-\mu Q}}{{\rm Tr}[e^{-\mu Q}]},    
\end{align}
the corresponding ensemble average being $\braket{\bullet}\equiv {\rm Tr}[\varrho_\mu(\bullet)]$. In a generic ratchet circuit, which lacks additional conservation laws, $\varrho_\mu$ is in fact the most general form of a Gibbs ensemble.

Owing to the lack of $\mathcal{P}$ symmetry in a general
quantum many-body spin ratchet with $s_{1}\ne s_{2}$, the
dynamical structure factor $S(x,t)$ is not symmetric under the spatial reflection, i.e., $S(x,t)\neq S(-x,t)$. Consequently, $S(x,t)$ will feature a finite hydrodynamic drift with velocity
\begin{equation}
\label{eq:general-drift-velocity}
v_{\rm d} \equiv \frac{1}{\chi}\lim_{t\to \infty}\int {\rm d}x\,\frac{x}{t}\,S(x,t),   
\end{equation}
normalized by the time-independent static spin susceptibility 
\begin{align}
\label{eq:static-spin-susceptibility}
    \chi\equiv\int{\rm d}x \, S(x,t).
\end{align}
To fully quantify the asymmetry, we will consider the asymptotic time-scaled centered moments of the dynamical structure factor defined as
\begin{align}
\label{eq:scaled-cumulant_susceptibility}
    S^{(n)}\equiv \lim_{t\to\infty}\int {\rm d}x \left(\frac{x}{t}-v_{\rm d}\right)^n S(x,t).
\end{align}
Note that the first centered moment trivially vanishes by definition, i.e., $S^{(1)}=0$. The second centered moment, however, is the Drude weight, $S^{(2)}=\mathcal{D}\geq 0$. It quantifies ballistic spreading of local density disturbances in the frame moving with the drift velocity $v_{\rm d}$. Higher centered moments $S^{(n)}$ quantify deviations from Gaussianity. We do not consider them explicitly herein.

\subsubsection{Hydrodynamics in an integrable ratchet}

On large space-time scales, the dynamical structure factor of an integrable ratchet can be accurately described by means of the generalized hydrodynamics~\cite{alvaredo2016,bertini2016}. 
Specifically, on the ballistic (Euler) hydrodynamic scale, characterized by large $x$ and $t$ with their ratio $x/t$ fixed, the dynamical structure factor $S(x,t)$ admits a mode resolution in terms of quasiparticle excitations. The latter are accessible within the thermodynamic Bethe ansatz (TBA) which describes the thermodynamic eigenstates of the model with a finite set of state functions per each mode---see Appendix~\ref{app:TBA} for information concerning the thermodynamic limit of the Bethe ansatz equations~\eqref{eq:bethe_equations}. Similarly to other integrable spin chains, the integrable spin ratchet features
quasiparticles (magnons) that undergo elastic scattering.
Due to attractive interaction, magnons can form bound states with $m$ quanta of magnetization, commonly referred to as the bare charge $q_m(\lambda)\equiv q_m=m$~\footnote{The bare charge $q_m$ of a bound state should not be confused with the on-site magnetization densities defined in Eq.~\eqref{eq:local_mag}, which always have the upper index indicating the odd or even bond.}.
Bare quasimomenta of such bound states are parametrized by the rapidity $\lambda \in \mathbb{R}$. Owing to mutual interaction, bare quantities of quasiparticles undergo nontrivial renormalization called \emph{dressing}. Dressing corresponds to a linear transformation depending on both the state occupation function and the scattering data. For instance, $q_m^{\rm dr}$ denotes the dressed magnetization of a bound state, as detailed out in Appendix~\ref{app:TBA-dressing}.

On ballistic hydrodynamic scale, the mode resolution of the structure factor takes the form of a weighted sum of delta peaks~\cite{denardis2022,doyon2017}, 
\begin{align}
\label{eq:dynamical_susceptibility}
    S(x,t)\simeq  \sum_{m=1}^{\infty}\int{\rm d}\lambda\, \delta \left(x-v^{\rm eff}_m(\lambda) t\right)\chi_m(\lambda) (q_m^{\rm dr})^2,
\end{align}
propagating with effective mode velocities~\footnote{Here the quasiparticles' bare quasienergy and quasimomentum, namely $\varepsilon_m(\lambda)$ and $p_m(\lambda)$, should not be confused with the eigenvalues~\eqref{eq:quasimomentum} and~\eqref{eq:quasienergy} of the lattice-shift operators~\cite{alvaredo2016,bertini2016}.} 
\begin{align}
\label{eq:effective_velocity}
    v^{\rm eff}_m(\lambda)=\frac{(\varepsilon'_m)^{\rm dr}(\lambda)}{(p'_m)^{\rm dr}(\lambda)},
\end{align}
where $f'(\lambda)\equiv\partial^{}_\lambda f(\lambda)$ denotes the derivative on the rapidity. In Eq.~\eqref{eq:dynamical_susceptibility} we have introduced mode susceptibilities
$\chi_m(\lambda)\equiv\rho^{\rm tot}_m(\lambda) n_m(\lambda)[1-n_m(\lambda)]$, in which $\rho^{\rm tot}_m(\lambda)$ denote the total densities of available states in the rapidity space, while $n_m(\lambda)\equiv \rho_{m}(\lambda)/\rho^{\rm tot}_{m}(\lambda)$ are the occupation fractions. 
The drift velocity~\eqref{eq:general-drift-velocity} now becomes
\begin{align}
\label{eq:drift_velocity}
v_{\rm d}=\frac{1}{\chi}\sum_{m=1}^\infty\int{\rm d}\lambda\chi_m(\lambda) v^{\rm eff}_m(\lambda)(q_m^{\rm dr})^2,
\end{align}
where the static spin susceptibility~\eqref{eq:static-spin-susceptibility} admits the following mode decomposition:
\begin{align}
\chi=\sum_{m=1}^{\infty}\int{\rm d}\lambda\, \chi_m(\lambda) (q_m^{\rm dr})^2.
\end{align}

In a system with an unbroken $\mathcal{P}$ symmetry, the occupancies $n_m(\lambda)$ and the total state densities $\rho_m^{{\rm tot}}(\lambda)$ entering the mode susceptibilities $\chi_m(\lambda)$ are even functions of the rapidity $\lambda$. On the other hand, $v_m^{\rm eff}(\lambda)$ is an odd function of $\lambda$ and, as a result, the drift velocity vanishes. This is the case in homogeneous circuits with $s_{1}=s_{2}$.
Instead, in ratchets with $s_{1}\ne s_{2}$, $\mathcal{P}$ symmetry is absent, and it is thus not surprising that we find a finite $v_{\rm d}$. Strikingly, however, we observe that $v_{d}$ is \emph{universal}. To corroborate this statement, we compute $v_{\rm d}$ explicitly in integrable spin ratchets, using the mode decomposition~\eqref{eq:drift_velocity}. In the Gibbs state~\eqref{eq:Gibbs-state}, the mode occupation functions $n_m$ become flat, i.e., they lose dependence on the rapidity $\lambda$, in turn simplifying the calculation. The latter involves an explicit solution of the infinite-temperature TBA equations of the spin-$s$ Heisenberg model which, to the best of our knowledge, has not been obtained before (see Appendix~\ref{app:TBA-solution}). The solution simplifies at half-filling, i.e., at $\mu=0$, where the right-hand side of Eq.~\eqref{eq:drift_velocity} can be evaluated explicitly {\em analytically}, yielding a remarkably simple expression
\begin{empheq}[box=\fbox]{align}
\label{eq:casimir-velocity}
    v_{\rm d}(\mu=0)=\frac{s_1(s_1+1)-s_2(s_2+1)}{s_1(s_1+1)+s_2(s_2+1)},
\end{empheq}
which notably depends only on the SU(2) Casimir invariants $s(s+1)$ of the local spin degrees of freedom. Note, moreover, that the above form of $v_{\rm d}$ reflects that of the effective velocity~\eqref{eq:effective_velocity} determined by the quasiparticles' quasienergies and quasimomenta. The latter are, respectively, a difference and a sum of the quasimomenta in the homogeneous Heisenberg chains with spins $s_1$ and $s_2$, similarly to the eigenvalues~\eqref{eq:quasienergy} and~\eqref{eq:quasimomentum} of the time-shift and space-shift operators. Specifically, they read $\varepsilon_m(\lambda)=[p^{(2s_1)}_m(\lambda_+)-p^{(2s_2)}_m(\lambda_-)]/2$ and $p_m(\lambda)=[p^{(2s_1)}_m(\lambda_+)+p^{(2s_2)}_m(\lambda_-)]/2$.

In the semiclassical limit $s_{1}, s_2 \to \infty$, the drift velocity of the symplectic ratchet \eqref{classical_map} reduces to a function of spin lengths $v_{\rm d} = (r_1^2 - r_2^2)/(r_1^2 + r_2^2)$, which we have verified by direct numerical simulations.

\subsubsection{Exact drift velocity in the noninteracting limit}

Recall that as $\tau\to\infty$, the integrable ratchet reduces to a simple brickwork circuit composed of permutation gates.
In this limit we can compute the drift velocity~\eqref{eq:casimir-velocity} exactly by evaluating the dynamical structure factor at coarse-grained integer coordinates $\ell\in\mathbb{Z}$, associated with pairs of lattice sites $(2\ell,2\ell+1)$, and at a discrete time $t=1$. Specifically, let us consider
\begin{align}
    S(\ell,1)\equiv\langle \mathbb{U}^{-1}q(\ell)\mathbb{U}q(0)\rangle^c=\langle q(\ell)\mathbb{U}q(0)\mathbb{U}^{-1}\rangle
\end{align}
at half-filling ($\mu=0$), where we have taken 
$q(\ell)\equiv q^{\rm (e)}_{2\ell}/2$, see Eq.~\eqref{eq:local_mag}. Since the time evolution is represented by a sequence of permutations, we observe that $\mathbb{U}q(0)\mathbb{U}^{-1}=(S^z_{1;3}+S^z_{2;-2})/2$, and hence the drift velocity reads simply
\begin{align}
    \label{eq:drift_vel_perm}
    v_{\rm d}(\mu\!=\!0)&=\frac{\sum_{\ell}\ell\langle q(\ell)\mathbb{U}q(0)\mathbb{U}^{-1}\rangle}{\sum_{\ell}\langle q(\ell)\mathbb{U}q(0)\mathbb{U}^{-1}\rangle}\notag\\[0.5em]
    &=\frac{\langle (S^z_1)^{2}\rangle-\langle(S^z_2)^{2}\rangle}{\langle (S^z_1)^{2}\rangle+\langle(S^z_2)^{2}\rangle}.
\end{align}
Using that, at zero magnetization density, one has $\langle (S^z)^2\rangle=s(s+1)/3$, we find precisely the Casimir-dependent drift velocity~\eqref{eq:casimir-velocity}. In fact, the simplicity of the circuit composed of permutation gates allows us to generalize this computation away from half-filling, i.e., for a general chemical potential $\mu\ne 0$. There, linear response relates the gradient of the coarse-grained magnetization profile $q(\ell,t)=\mathbb{U}^{-t}(q^{\rm (e)}_{2\ell}/2)\mathbb{U}^t$ to the dynamical charge susceptibility, $S(\ell,t)\propto\lim_{\delta\mu\to0}(\delta\mu)^{-1}\langle q(\ell+1,t)-q(\ell,t)\rangle_{\delta\mu}$~\cite{ljubotina2019-2}. Here, $\langle\bullet\rangle_{\delta\mu}$ denotes the expectation value in a bipartite state with chemical potentials $\mu_{\rm L/R}=\mu\pm\delta\mu$ on the left/right-hand side of the system. Exploiting this relation, we obtain
\begin{empheq}[box=\fbox]{align}
\label{eq:drift_general-mu}
    v_{\rm d}(\mu)=\frac{\partial_\mu \langle S_1^z\rangle-\partial_\mu \langle S_2^z\rangle}{\partial_\mu \langle S_1^z\rangle+\partial_\mu \langle S_2^z\rangle},
\end{empheq}
where
\begin{align}
\label{eq:eq:drift_general-mu_ingredient}
    \partial_\mu \langle S^z\rangle\!=\!\left\{\!\left(s\!+\!\tfrac{1}{2}\right)\!\mathrm{csch}\!\left(\mu\!\left[s\!+\!\tfrac{1}{2}\right]\right)\!\right\}^2\!-\!\left[\tfrac{1}{2}\mathrm{csch}\!\left(\tfrac{\mu}{2}\right)\right]^2\!.
\end{align}
This result remains valid for any value of $\tau>0$, which is corroborated by a numerical evaluation of the exact hydrodynamic formula~\eqref{eq:drift_velocity} away from half-filling. For instance, setting $s_1=1$, $s_2=1/2$, and $\mu=5/4$, with arbitrary $\tau$, we obtain $v_{\rm d}\approx 0.326338$ from Eq.~\eqref{eq:drift_general-mu} and $v_{\rm d}\approx 0.32633$ from the GHD expression~\eqref{eq:drift_velocity}. In the numerical evaluation of the latter, we have integrated over rapidities $\lambda\in [-8\!\times\! 10^3, 8\!\times \!10^3]$ and truncated the sum over modes to $m_{\rm max}=500$ terms.

The fact that the drift velocity takes a simple universal form~\eqref{eq:casimir-velocity} (or Eq.~\eqref{eq:drift_general-mu} for $\mu\ne 0$) suggests that it might result purely from the spin-exchange operator entering the local unitary map $U$ of a quantum many-body spin ratchet. As shown in the following, the drift velocity formulae~\eqref{eq:casimir-velocity} and~\eqref{eq:drift_general-mu} indeed continue to hold even in generic (i.e., chaotic) spin ratchets. We stress again that the drift in the ratchet circuit is a direct consequence of the broken parity symmetry. Curiously, a similar effect has been reported in the context of entanglement spreading in cellular automata defined on staggered lattices made out of local Hilbert spaces of different dimensions~\cite{piroli2022}. There, a background velocity depending only on the logarithms of the local Hilbert space dimensions has been observed.

\begin{figure*}[tb!]
    \centering
    \includegraphics[width=0.95\textwidth]{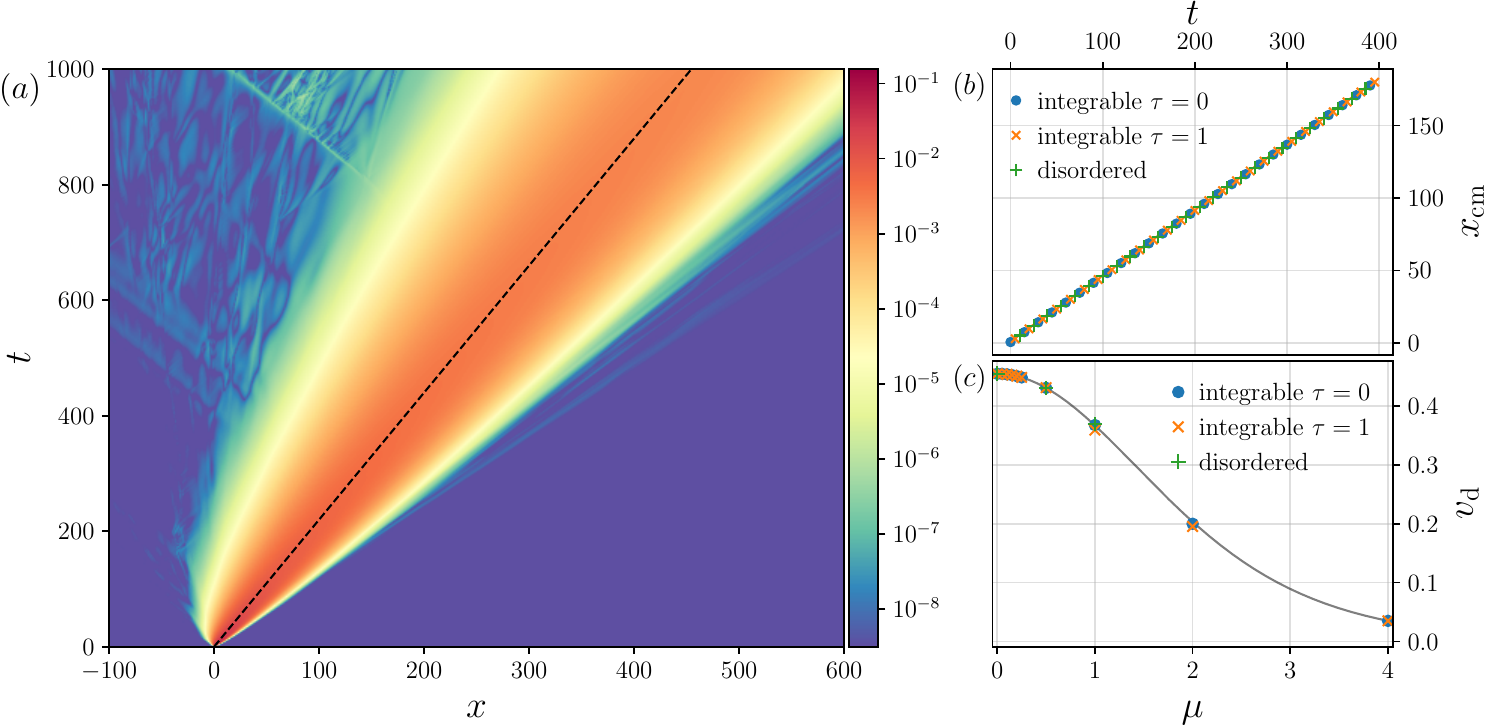}
    \vspace{1ex}
    \caption{(a) Dynamical charge susceptibility $S(x, t)$ at $\tau=1$ and $\mu=0$, for a circuit with $s_1=1$ and $s_2=1/2$. The dashed black line corresponds to the drift velocity $v_{\rm d}(\mu=0)=5/11$, computed from the center of mass ($x_{\rm cm}$) of $S(x,t)$ at each time $t$, as detailed in panel (b). (c) Drift velocity $v_{\rm d}$ (for $s_1=1, s_2=1/2$) estimated from tensor  network simulations of an integrable ratchet with $\tau=0$ (blue circles), $\tau=1$ (orange crosses), and of a disordered system (green pluses), compared against the analytical prediction \eqref{eq:drift_general-mu}. We observe excellent agreement for all $\mu$ in both integrable and non-integrable circuits.}
    \label{fig:Sxt}
\end{figure*}

\begin{figure}[b!]
    \centering
    \includegraphics[width=0.95\linewidth]{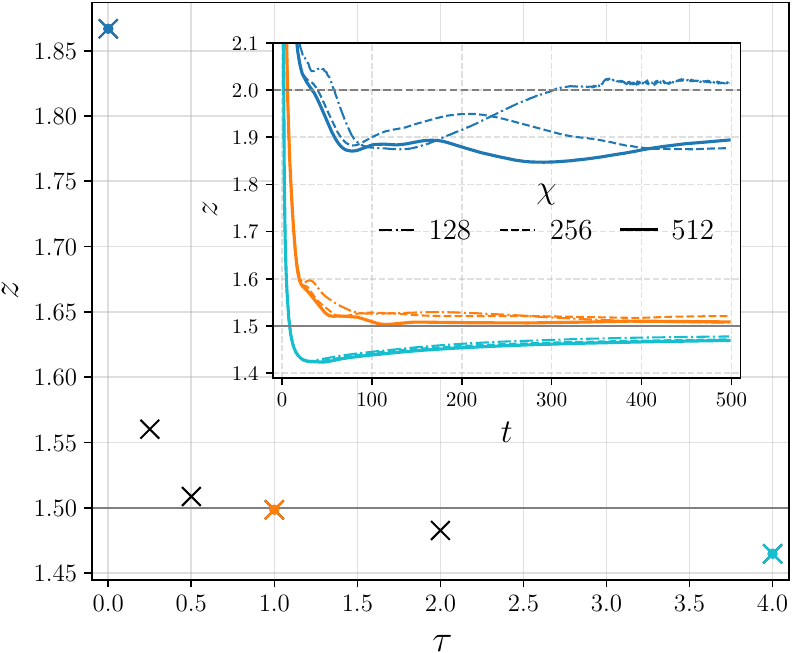}
    \vspace{1ex}
    \caption{Main figure: dynamical exponent $z$ for various unitary-gate parameters $\tau$ at half-filling ($\mu=0$), extracted from  tensor-network simulations with bond dimension $\chi$. Inset: convergence of $z$ with time $t$. The results indicate superdiffusive scaling compatible with $z=3/2$ for all $\tau >0$, in line with expectations for an integrable model with SU(2) symmetry. At the exceptional point $\tau=0$, the diffusion constant diverges logarithmically with time $t$, while numerical results indicate an approximate diffusive scaling on accessible time scales.}
    \label{fig:zexp}
\end{figure}

\subsubsection{Universality of drift velocity and spreading of correlations}

To verify the conjectured universal formula for the drift velocity we performed large-scale tensor-network  simulations on integrable and disordered (non-integrable) ratchet circuits. The dynamical charge susceptibility $S(x,t)$, shown in Fig.~\hyperref[fig:Sxt]{2a} for the integrable ratchet with $s_1=1$, $s_2=1/2$, $\tau=1$, and at $\mu=0$, features a nonzero first moment corresponding to a drift velocity $v_{\rm d}(\mu=0)=5/11$. The drift velocities extracted from numerical simulations, shown in Fig.~\hyperref[fig:Sxt]{2b} for different integrable and nonintegrable instances, all coincide with the GHD results from Eq.~\eqref{eq:casimir-velocity}. 
Likewise, the drift velocities away from half-filling (cf. Fig.~\hyperref[fig:Sxt]{2c}) are in excellent agreement with the analytic formula derived in the noninteracting limit, and reported in Eqs.~\eqref{eq:drift_general-mu} and~\eqref{eq:eq:drift_general-mu_ingredient}.

Finally, to extract the algebraic dynamical exponent $z$, defined via the asymptotic growth of the second moment of $S(x,t)$,
\begin{equation}
    \sigma^{2}(t)\equiv \int {\rm d} x (x-v_{\rm d}t)^{2}\,S(x,t) \simeq t^{2/z},
\end{equation}
we also investigate the dynamics in the moving (i.e., center of mass) frame. In integrable ratchets, we generically expect ballistic growth with exponent $z=1$.
This is however the case only in generic states, i.e.,
away from half-filling. At $\mu=0$, the SU(2) symmetry of the state gets restored, affecting the type of spin transport in a profound way. 
Specifically, for $\tau>0$ we observe the anticipated fractional dynamical exponent $z=3/2$, in line with the general predictions for integrable models invariant under nonabelian symmetries---see Fig.~\ref{fig:zexp}.

The special point $\tau=0$ is however exceptional and exhibits a qualitatively different behavior. This fact can already be recognized at the level of bare quasiparticle dispersion relations. Accordingly, by repeating the scaling analysis of Refs.~\cite{gopalakrishnan2019,denardis2020}, we infer an anomalous type of diffusion, with a singular diffusion constant $D$ \emph{diverging} logarithmically with time, $D\sim \log t$. Such a law has already been observed previously in integrable spin models~\cite{denardis2020}. We note that such a mild divergence cannot be reliably resolved with accessible numerics (see the inset in Fig.~\ref{fig:zexp}) which instead hints at a normal diffusive scaling ($z=2$).

\subsection{Large-scale current fluctuations}
\label{sec:fluctuations}

As an alternative dynamical probe of spin transport we now 
consider the scaling of fluctuations of the time-integrated current density [see Eq.~\eqref{eq:charge-current}] across a site (say, at $x=0$) in the middle of an extended system,
\begin{align}
\label{eq:time-integrated_current}
    \mathfrak{J}_t=\int_{0}^t{\rm d}t' \left(j(0,t')-\langle j\rangle\right).
\end{align}
Here, we have subtracted the finite average value of the background current inherent to our ratchet systems ($\langle\bullet\rangle$ denotes the ensemble average)~\footnote{This is allowed, since the current in the continuity equation~\eqref{eq:charge-current} is defined up to an additive constant shift.}.
We again study the transport in maximum-entropy stationary states described by the density matrix~\eqref{eq:Gibbs-state}. For concreteness we focus on integrable ratchet circuits, where large-scale fluctuations can be characterized exactly.

The time-integrated current $\mathfrak{J}_t$ is a macroscopic fluctuating variable, whose values $\mathfrak{J}$ are distributed according to a probability distribution $\mathbb{P}(\mathfrak{J}|t)$.
Typical values of $\mathfrak{J}$ are of the order
$\mathfrak{J}\sim O(t^{1/{2z}})$, where $z$ is the dynamical exponent governing the decay of the density and current two-point functions. Generically, the distribution of typical values tends to a Gaussian at late times, i.e., it complies with the central-limit behavior.
On the other hand, in certain systems featuring dynamical criticality~\cite{krajnik2022,krajnik2024}, one finds it converging to a universal non-Gaussian asymptotic distribution. In the integrable ratchets, such a critical behavior is expected in the unbiased ensemble at $\mu=0$.

Here, we consider $\mu>0$, and instead examine the structure of large fluctuations. Particularly, we are interested in the fluctuations of the time-integrated current on the largest ballistic scale, with $\mathfrak{J} \simeq \mathfrak{j}\,t$ at large $t$ (for $0<\mathfrak{j}<\infty$). For asymptotically large times $t$, rare events are expected to obey a large deviation principle
\begin{align}
    \mathbb{P}(\mathfrak{J}=\mathfrak{j}t|t)\asymp e^{-t I(\mathfrak{j})},
\end{align}
where $\asymp$ signifies the asymptotic logarithmic equivalence for large $t$ and $I(\mathfrak{j})$ is the large-deviation rate function. If the rate function $I(\mathfrak{j})$ is differentiable, the G\"{a}rtner-Ellis theorem states that its Legendre-Fenchel transform $F(\zeta)\equiv \max_{\mathfrak{j}}[\zeta \mathfrak{j}-I{(\mathfrak{j})}]$ is the scaled cumulant generating function (SCGF) of the time-integrated current~\eqref{eq:time-integrated_current}~\cite{touchette2009},
\begin{align}
\label{eq:dynamical-free-energy}
    F(\zeta)=\lim_{t\to\infty}\frac{1}{t}\log\langle e^{\zeta \mathfrak{J}_t}\rangle.
\end{align}
Provided that a certain regularity condition is satisfied, the derivatives of the SCGF at $\zeta=0$ correspond to the scaled cumulants of the time-integrated current~\cite{bryc1993}, i.e., $({\rm d}^n/{\rm d}\zeta^n) F(\zeta)|_{\zeta=0} \equiv c^{(\rm sc)}_n$, where
\begin{align}
\label{eq:scaled-cumulants}
    c^{(\rm sc)}_n\equiv\lim_{t\to\infty}\frac{1}{t}\langle \mathfrak{J}_t^n\rangle^{c}.
\end{align}
We stress that such a regular behaviour of the SCGF is not guaranteed. A notable counterexample are models exhibiting dynamical criticality~\cite{krajnik2024,krajnik2024-2,rosenberg2024}. The latter are characterized by cumulants $c_n(t)=\langle\mathfrak{J}^n_t\rangle^c$ that do not all scale with the same power of $t$. In such a case, the corresponding scaled cumulants~\eqref{eq:scaled-cumulants} are ill-defined.

The integrable spin ratchet circuit considered here exhibits dynamical criticality at $\mu=0$. In the following,
we instead specialize to the regular regime $\mu\ne 0$ where all scaled cumulants~\eqref{eq:scaled-cumulants} exist and are given by the derivatives of the SCGF at $\zeta=0$. In this case, $c^{(\rm sc)}_{n}$ are accessible within the GHD.
In particular, the first scaled cumulant is trivially zero, $c_1^{(\rm sc)}=0$. This is simply due to the subtracted average current $\langle j\rangle$ in Eq.~\eqref{eq:time-integrated_current} which only changes the linear slope of the SCGF~\eqref{eq:dynamical-free-energy} (thereby removing the effect of the broken $\mathcal{P}$ symmetry), whereas all higher scaled cumulants remain unaffected. The second scaled cumulant, also known in the literature as the Drude self-weight~\cite{doyon2017}, denoted by $\mathcal{D}_{\rm self}$, corresponds to the growth rate of the first absolute moment of the dynamical structure factor,
\begin{align}
\label{eq:self-Drude}
    c_2^{(\rm sc)}=\mathcal{D}_{\rm self}\equiv \lim_{t\to\infty}\frac{1}{t}\int{\rm d}x |x| S(x,t).
\end{align}
It admits the following mode resolution,
\begin{align}
\label{eq:GHD_self-Drude}
    c_2^{(\rm sc)}=\sum_{m=1}^\infty \int{\rm d}\lambda |v^{\rm eff}_m(\lambda)|\chi_m(\lambda) (q_m^{\rm dr})^2,
\end{align}
which can be evaluated numerically. We have also compared it to direct numerical calculation of the absolute moment~\eqref{eq:self-Drude} using our tensor-network (TN) simulations. For example, setting $s_1=1$, $s_2=1/2$, $\mu=1$, and $\tau=1$, the GHD formula~\eqref{eq:GHD_self-Drude} yields $c_2^{(\rm sc)}\approx 0.1407$ (obtained with the cutoff $m_{\rm max}=20$ and using integration over the compact rapidity domain $\lambda\in [-5\!\times\! 10^2,5\!\times\! 10^2]$). On the other hand, the TN simulation results in $c_{2}^{(\rm sc)}\approx 0.14(0)$. Similarly, in the limit $\tau\to 0$ and with other parameters unchanged, we have $c_2^{(\rm sc)}\approx 0.1262$ from the GHD and $c_{2}^{(\rm sc)} \approx 0.126(1)$ from the TN simulation.

The hydrodynamic mode expansion of higher cumulants can be systematically derived from the ballistic fluctuation theory~\cite{myers2020,doyon2023}, or by using diagrammatic techniques~\cite{vu2020}. For example, the third scaled cumulant can be computed as a sum of two terms (see, e.g., Refs.~\cite{myers2020,vu2020,doyon2023})
\begin{align}
\label{eq:third-scaled-cumulant}
    c^{\rm (sc)}_3 = c^{({\rm sc})}_{3;1} + c^{({\rm sc})}_{3;2},
\end{align}
with hydrodynamic mode resolutions
\begin{align}
\begin{aligned}
\label{eq:third-scaled-cumulant-contributions}
    c^{\rm (sc)}_{3;1} &=\sum_{m=1}^\infty \int \frac{{\rm d}\lambda}{2\pi}w^{(3)}_m(\varepsilon'_m)^{\rm dr}(\lambda) (q_m^{\rm dr})^3,\\
    c^{\rm (sc)}_{3;2} &=3 \sum_{m=1}^\infty \int \frac{{\rm d}\lambda}{2\pi}w^{(2)}_m \sigma^{}_m(\lambda)\gamma^{}_{m}(\lambda)(\varepsilon'_m)^{\rm dr}(\lambda) q_m^{\rm dr}.
\end{aligned}
\end{align}
Here, 
$\sigma^{}_m(\lambda)\equiv{\rm sgn}[v^{\rm eff}_m(\lambda)]={\rm sgn}[(\varepsilon'_m)^{\rm dr}(\lambda)]$ is the sign of the effective velocity,
\begin{align}
    w_m^{(k)}=\sum_{r=1}^\infty (-1)^{r-1} r^{k-1} \left(\frac{n_m}{1-n_m}\right)^r
\end{align}
are statistical weights, and we have introduced
\begin{equation}
\label{eq:screened-function}
 \gamma^{}_m(\lambda)\equiv -\left[(1-n^{}_m)\sigma^{}_m (q_m^{\rm dr})^2\right]^{\rm scr}(\lambda),  
\end{equation}
where the screening operation is defined as $f^{\rm scr}(\lambda)\equiv f(\lambda)-f^{\rm dr}(\lambda)$ (see Appendix~\ref{app:TBA-dressing}).

\subsubsection{Breakdown of the Gallavotti--Cohen relation}

In dynamical systems with time-reversal symmetry, the scaled cumulant generating function obeys the Gallavotti--Cohen relation (GCR) \cite{bernard2013,doyon2023}. In the absence of thermodynamic forces it reads
\begin{align}
\label{eq:gallavotti-cohen}
    F(\zeta)=F(-\zeta),
\end{align}
and it indicates a lack of directionality: the probability of observing a large time-integrated current $\mathfrak{J}\sim O(t)$ does not depend on its direction. As established in Ref.~\cite{doyon2023}, the time-reversal symmetry of the Euler-scale hydrodynamics is sufficient for the validity of  GCR. It nonetheless remains an open question whether it is also a necessary condition. Our aim here is to address this question in the present context of quantum spin ratchets
which, while lacking $\mathcal{T}$ symmetry, obey the $\mathcal{P}\mathcal{T}$ symmetry. Specifically, we take a look at the third scaled cumulant and compute it using Eqs.~\eqref{eq:third-scaled-cumulant} and~\eqref{eq:third-scaled-cumulant-contributions}. A non-zero value of $c^{\rm (sc)}_3$ implies the violation of GCR.

By numerically evaluating the hydrodynamic mode expansions in Eq.~\eqref{eq:third-scaled-cumulant-contributions}, we find that $c^{\rm (sc)}_3$ is indeed nonzero. For example, setting $s_1=3/2$, $s_2=1/2$, with $\tau=1$ and $\mu=1$, we obtain $c_{3;1}^{\rm (sc)}\approx 0.6328$ and $c_{3;2}^{\rm (sc)}\approx -0.343(1)$ (integrating over the rapidity domain $\lambda\in[-5\!\times\! 10^2,5\!\times\! 10^2]$ and taking the cutoff $m_{\rm max}=20$). Combining both terms yields a nonzero third scaled cumulant. Violation of the GCR in a (grand-canonical) Gibbs ensemble~\eqref{eq:Gibbs-state} indicates that, in ratchets, such stationary maximum-entropy ensembles do not describe a thermodynamic equilibrium.

\subsubsection{Generalized fluctuation symmetry}

In turns out that integrable ratchet circuits nevertheless exhibit a generalized fluctuation symmetry relation. Owing to the absence of $\mathcal{P}$ symmetry, we can establish a relation between the current fluctuations in the ratchet and its spatially reflected counterpart, in which the alternating spins $s_1$ and $s_2$ are exchanged. Specifically, denoting the SCGF of the ratchet circuit with alternating spins $s_1$ and $s_2$ by $F^{(s_1,s_2)}(\zeta)$, we have
\begin{empheq}[box=\fbox]{align}
\label{eq:generalized-fluctuations-symmetry}
    F^{(s_1,s_2)}(\zeta)=F^{(s_2,s_1)}(-\zeta).
\end{empheq}
This can be inferred from the expressions for the scaled cumulants, which have all been conjectured to admit a diagrammatic expansion~\cite{vu2020}. In those expressions, the only functions that depend on spins $s_1$ and $s_2$ are the dressed derivatives of the quasiparticle dispersion relations
$\varepsilon_m(\lambda)\equiv\varepsilon_m^{(s_1,s_2)}(\lambda)$, and their signs $\sigma_m(\lambda)\equiv\sigma_m^{(s_1,s_2)}(\lambda)$. Here, we have introduced the upper index to denote the alternating spins in the ratchet circuit. The derivatives of the bare quasienergies satisfy the symmetry
\begin{align}
\label{eq:quasienergies-symmetry}
    \partial^{}_\lambda\varepsilon_m^{(s_1,s_2)}(\lambda)=-\partial^{}_\lambda\varepsilon_m^{(s_2,s_1)}(-\lambda),
\end{align}
which can be verified already for the derivative of the single-magnon quasienergy~\eqref{eq:quasienergy}, the latter written in terms of the single-magnon quasimomenta~\eqref{eq:single-particle-momentum}. Since the dressing operation maps as $[-f(-\lambda)]\mapsto [-f^{\rm dr}(-\lambda)]$ and does not itself depend on spins $s_{1}$ and $s_2$ (see Appendix~\ref{app:TBA-dressing}), it preserves the relation in Eq.~\eqref{eq:quasienergies-symmetry}. The generalized fluctuation symmetry~\eqref{eq:generalized-fluctuations-symmetry} then readily follows from the following observations:
\begin{enumerate}
    \item The hydrodynamic formulae for all scaled cumulants involve one dressed derivative of the dispersion under the integral over the rapidity $\lambda$.
    \item The odd (even) cumulants additionally involve  an even (resp. odd) number of signs $\sigma^{(s_1,s_2)}_m(\lambda)$ [see, e.g., Eqs.~\eqref{eq:GHD_self-Drude},~\eqref{eq:third-scaled-cumulant}, and~\eqref{eq:screened-function}].
    \item Changing the integration variable as $(-\lambda)\mapsto \lambda$ does not affect the hydrodynamic formulae for the scaled cumulants.
\end{enumerate}

\section{Discussion}
\label{sec:discussion}

With the aim to characterize the dynamical effects of broken space-time symmetries, we have introduced and studied a family of quantum circuits made out of SU(2)-symmetric unitary gates with an adjustable free parameter, acting on spins of unequal sizes. While the constructed models explicitly break the space-reflection and time-reversal symmetries, they preserve the combined $\mathcal{P}\mathcal{T}$ symmetry. Depending on the choice of parameters, the circuit can be made ergodic or integrable. The latter, in particular, generalizes the integrable Trotterization of the isotropic Heisenberg spin-$1/2$ chain~\cite{vanicat2018}. The breaking of the $\mathcal{P}$ symmetry induces a chiral spin dynamics, due to which we can view our circuit as a many-body analogue of a quantum ratchet. We outline how such ratchets can be experimentally implemented by encoding higher spins using multilevel trapped ions~\cite{Ringbauer_2022}.

Quantum spin ratchets have two defining dynamical properties. Firstly, they  exhibit a drift in the dynamical structure factor, which we have quantified by analytically deriving a simple universal formula for the drift velocity. The latter depends only on the size of spins and on the magnetization density in the initial Gibbs ensemble, but not on the microscopic details of the local unitary gates. In the integrable circuits, we have retrieved the formula in a fully analytic manner, in the scope of the generalized hydrodynamics, as a result of a nontrivial resummation over the spectrum of quasiparticle excitations. This required a closed-form analytic solution of the thermodynamic Bethe ansatz equations of the Heisenberg spin-$s$ chain, which, as far as we are aware, has not been reported previously.

The second key property of quantum many-body spin ratchets concerns the anomalous nature of the macroscopic current fluctuations in Gibbs states,
which is manifested at the level of the full counting statistics of the time-integrated spin current density. In the absence of thermodynamic forces, probabilities of large current fluctuations in time-reversal symmetric systems do not depend on the direction of the flow, as encapsulated by the Gallavotti--Cohen fluctuation symmetry. In stark contrast, in quantum many-body spin ratchets the fluctuation symmetry no longer holds in Gibbs states.
This consequently indicates that such states do not represent thermodynamic equilibrium states of many-body ratchets. Instead, we demonstrate that in quantum many-body ratchets the Gallavotti--Cohen relation is superseded by a generalized fluctuations symmetry: the large current fluctuations in one direction are connected to fluctuations in the opposite direction, in a spatially reflected system.

Our study of the dynamics in quantum many-body spin ratchets opens up several interesting research directions. Particularly, there remain several questions concerning the universality of the drift velocity:
\begin{itemize}
    \item[--] As demonstrated, the drift velocity is insensitive to integrability, so far as the functional form of the local unitary gate is preserved. We remind that breaking of integrability has been achieved through the choice of gate parameters. It however remains unclear 
    whether relaxing the functional form of the unitary gates (while preserving the SU(2) symmetry) can have any impact. We leave this aspect to future studies. 
    \item[--] A similar type of drift has been reported to underlie the coarse-grained entanglement dynamics in a staggered circuit studied in Ref.~\cite{piroli2022}, albeit the precise form of the drift velocity therein differs from ours. For a more comprehensive understanding, it would be worthwhile investigating various other realizations of unitary circuits with a spatial staggered structure. An example is the recently proposed circuit whose unitary gates encode the two-body scattering of particles with different masses in a supersymmetric quantum field theory~\cite{richelli2024}.
\end{itemize}

The second part of our work stimulates fundamental questions pertaining to the role of the space-time symmetries on macroscopic dynamical phenomena and, more specifically, the properties of atypical current fluctuations in maximum-entropy stationary states:
\begin{itemize}
    \item[--] It is important to determine whether the established generalized fluctuation relation~\eqref{eq:generalized-fluctuations-symmetry} requires both $\mathcal{P}\mathcal{T}$ and
    charge-conjugation symmetry $\mathcal{C}$ independently,
    or there is perhaps a more general fluctuation relation hinging only on the $\mathcal{C}\mathcal{P}\mathcal{T}$ invariance.
    \item[--] While it appears plausible that $\mathcal{T}$-symmetry is not only sufficient but also necessary for the validity of the Gallavotti--Cohen relation, this still remains formally unresolved.
    Investigation of spontaneous-current fluctuations~\cite{eliashberg1983,tavger1986,blount1988,halperin1989,simon1992,roy2018,madsen2021,kobayashi2022} could provide a valuable insight on this question.
\end{itemize}

\begin{acknowledgments}
The authors thank Denis Bernard, J\'er\^ome Dubail, Hosho Katsura, Kareljan Schoutens, and Alberto Zorzato for stimulating discussions. 

This work has been supported by:
\begin{itemize}
    \item[--] Slovenian Research Agency (ARIS) under grants
    N1-0219 (T.P., L.Z.), N1-0334 (T.P., L.Z.), N1-0243 (E.I.), and under research program P1-0402 (E.I., T.P., L.Z.).
    \item[--] European Research Council (ERC) under Consolidator Grant No.~771536 -- NEMO (L.Z.), Advanced grant No.~101096208 -- QUEST (T.P., L.Z.), and Starting Grant. No.~850899 -- NEQuM (M.L.).
    \item[--] Simons Foundation under Simons Junior Fellowship grant No.~1141511 (\v{Z}.~K.).
\end{itemize}

M.~L. acknowledges the hospitality of the Aspen Center for Physics, which is supported by National Science Foundation grant PHY-2210452. 

Numerical simulations were performed using the ITensor library~\cite{ITensor}. 

\end{acknowledgments}

\appendix

\section{Embedding of a quantum ratchet}
\label{app:embedding}

Here, we show how ratchet circuits with $s_1\ne s_2$ can be realized in terms of quantum unitary gates acting on a homogeneous chain of spin-$1/2$ particles or qubits with local Hilbert space $\mathcal{H}_{1/2} \cong \mathbb{C}^{2}$. This is achieved by encoding higher spins into multi-qubit spaces.

The first step is to embed a local spin-$s_j$ space $\mathcal{H}_{s_j}$ into a space of $n_j$ qubits using embedding maps $\Omega_j:\mathcal{H}_{s_j}\xhookrightarrow{}\mathcal{H}^{\otimes n_j}_{1/2}=\mathcal{V}_{d_j}\oplus \overline{\mathcal{V}}_{d_j}$,
\begin{align}
\label{eq:app_embedding}
    \Omega_j=\sum_{m=-s_j}^{s_j} \ket{\alpha_m}\!\bra{m},
\end{align}
where $\ket{\alpha_m}$ form some orthonormal basis of a subspace $\mathcal{V}_{d_j}\cong \mathcal{H}_{s_j}$ of dimension $d_j=2s_j+1\le 2^{n_j}$.

Introducing $\Omega_{1,2}\equiv\Omega_1\otimes\Omega_2$, we next define a family of quantum gates $V_{\rm emb}\in{\rm End}[\mathcal{H}^{\otimes (n_1+n_2)}_{1/2}]$,
\begin{align}
    V_{\rm emb}=\Omega^{}_{1,2}V\Omega_{1,2}^\dagger+\left(\mathbbm{1}-\Omega^{}_{1,2}\Omega^\dagger_{1,2}\right)W,
    \label{eq:app_embedded-V}
\end{align}
where $W\in{\rm End}[\mathcal{H}^{\otimes (n_1+n_2)}_{1/2}]$ is an arbitrary unitary gate satisfying $[W,\mathbbm{1}-\Omega_{1,2}^{}\Omega_{1,2}^\dagger]=0$, and $\mathbbm{1}$ denotes an identity on $\mathcal{H}^{\otimes (n_1+n_2)}_{1/2}$. The first term in Eq.~\eqref{eq:app_embedded-V} represents embedding of $V$ into the product space $\mathcal{V}_{d_1}\otimes\mathcal{V}_{d_2}$, whereas the second term encodes an arbitrary unitary dynamics on its complement $\mathcal{H}^{(n_1+n_2)}_{1/2}\setminus (\mathcal{V}_{d_1}\otimes\mathcal{V}_{d_2})$, onto which $\mathbbm{1}-\Omega_{1,2}^{}\Omega_{1,2}^\dagger$ projects. Using that $\Omega^\dagger_j\Omega^{}_j=\mathbbm{1}_{\mathcal{H}_{s_j}}$ is an identity on $\mathcal{H}_{s_j}$, we deduce that
$V_{\rm emb}$ is unitary,
\begin{align}
    V^{}_{\rm emb}V_{\rm emb}^\dagger=\mathbbm{1}.
\end{align}

Finally, we require a permutation $P^{(n_1\leftrightarrow n_2)}$ that swaps the first $n_1$ qubits with the last $n_2$-ones, and which can be efficiently encoded as a sequence of pairwise SWAP gates. Applying the permutation after $V_{\rm emb}$, we obtain a unitary gate $U_{\rm emb}= P^{(n_1\leftrightarrow n_2)}V_{\rm emb}$ which encodes the local gate $U:\mathcal{H}_{s_1}\otimes\mathcal{H}_{s_2}\to \mathcal{H}_{s_2}\otimes\mathcal{H}_{s_1}$ of a quantum ratchet as an operator on $\mathcal{H}_{1/2}^{\otimes(n_1+n_2)}$. To form the circuit, the unitary gates $U_{\rm emb}$ must be arranged in a brickwork fashion, with each consecutive layer shifted $n_2$ sites to the right relative to the previous one.

\begin{definition}
To exemplify the construction, we explicitly work out the simplest case with $s_1=1$ and $s_2=1/2$, by embedding $U$ as an operator acting on the Hilbert space $\mathcal{H}^{\otimes 3}_{1/2}$ of three qubits. In particular, we embed the spin-$1$ Hilbert space into the triplet subspace of the first two qubits using $\Omega_1:\mathcal{H}_{1}\xhookrightarrow{}(\mathcal{H}_{1/2})^{\otimes 2}$ [cf. \eqref{eq:app_embedding}], given by
\begin{align}
    \Omega_1=\ket{\ua\ua}\!\bra{1}+\frac{1}{\sqrt{2}}\left(\ket{\ua\da}\!+\!\ket{\da\ua}\right)\!\bra{0}+\ket{\da\da}\!\bra{-1}.
\end{align}
The second spin, $s_2=1/2$, is instead identified with the third qubit---the corresponding embedding $\Omega_2:\mathcal{H}_{1/2}\xhookrightarrow{}\mathcal{H}_{1/2}$ is trivial, i.e., $\Omega_2=\mathbbm{1}_{\mathcal{H}_{1/2}}$.

Note that, since $\Omega_1\Omega_1^\dagger$ projects onto a triplet subspace of $\mathcal{H}_{1/2}^{\otimes 2}$, we have
\begin{align}
\label{eq:app_projector}
    \mathbbm{1}\!-\!\Omega^{}_{1,2}\Omega^\dagger_{1,2}=\operatorname{p}^{\rm (s)}\!\otimes\mathbbm{1}_{\mathcal{H}_{1/2}},
\end{align}
where 
\begin{align}
    \operatorname{p}^{\rm (s)}&=\frac{1}{2}\left(\ket{\uparrow\downarrow}\!-\!\ket{\downarrow\uparrow}\right)\!\left(\bra{\uparrow\downarrow}\!-\!\bra{\downarrow\uparrow}\right)
\end{align}
projects onto a one-dimensional subspace associated with a spin singlet. For a unitary operator $W$ commuting with the projector~\eqref{eq:app_projector}, there exists a unitary $2\times 2$ matrix $Z\in{\rm End}(\mathcal{H}_{1/2})$, such that $(\operatorname{p}^{\rm (s)}\!\otimes\mathbbm{1}_{\mathcal{H}_{1/2}})W=\operatorname{p}^{\rm (s)}\!\otimes\, Z$. The embedded gate~\eqref{eq:app_embedded-V} then reads
\begin{align}
 V_{\rm emb}=(\Omega^{}_1\otimes\mathbbm{1}_{\mathcal{H}_{1/2}})V(\Omega^\dagger_1\otimes\mathbbm{1}_{\mathcal{H}_{1/2}})+\operatorname{p}^{\rm (s)}\!\otimes\, Z.
\end{align}
For simplicity, we will set $Z=\mathbbm{1}_{\mathcal{H}_{1/2}}$ in the following.

Lastly, the permutation $P^{1,1/2}$ is embedded as $P^{(2\leftrightarrow 1)}={\rm SWAP}_{1,2}\,{\rm SWAP}_{2,3}$, interchanging the first two with the last qubit.
In the computational basis of the three-qubit Hilbert space, the embedded unitary gate thus reads
\begin{widetext}
\begin{align}
    U_{\rm emb}=P^{(2\leftrightarrow 1)}V_{\rm emb}=
    \begin{pmatrix}
     1 & 0 & 0 & 0 & 0 & 0 & 0 & 0 \\
     0 & \frac{2 i}{2 \tau +3 i} & \frac{2 (\tau +i)}{2 \tau +3 i} & 0 & -\frac{i}{2 \tau +3 i} & 0 & 0 & 0 \\
     0 & \frac{2 i}{2 \tau +3 i} & -\frac{i}{2 \tau +3 i} & 0 & \frac{2 (\tau +i)}{2 \tau +3 i} & 0 & 0 & 0 \\
     0 & 0 & 0 & \frac{2 i}{2 \tau +3 i} & 0 & \frac{2 i}{2 \tau +3 i} & \frac{2\tau- i}{2 \tau +3 i} & 0 \\
     0 & \frac{2\tau- i}{2 \tau +3 i} & \frac{2 i}{2 \tau +3 i} & 0 & \frac{2 i}{2 \tau +3 i} & 0 & 0 & 0 \\
     0 & 0 & 0 & \frac{2 (\tau +i)}{2 \tau +3 i} & 0 & -\frac{i}{2 \tau +3 i} & \frac{2 i}{2 \tau +3 i} & 0 \\
     0 & 0 & 0 & -\frac{i}{2 \tau +3 i} & 0 & \frac{2 (\tau +i)}{2 \tau +3 i} & \frac{2 i}{2 \tau +3 i} & 0 \\
     0 & 0 & 0 & 0 & 0 & 0 & 0 & 1 \\
    \end{pmatrix}.
    \end{align}
\end{widetext}
\end{definition}

We note that instead of using an embedding into a multi-qubit space, one could have alternatively employed a two-fold copy of the larger spin space, namely, one could have used an embedding $\mathcal{H}_{s_1}\otimes\mathcal{H}_{s_2}\xhookrightarrow{}\mathcal{H}_{s_{\rm max}}\otimes\mathcal{H}_{s_{\rm max}}$, where $s_{\rm max}={\rm max}(s_1,s_2)$. Such an embedding could, for instance, be utilized in a quantum processor based on trapped ions~\cite{Ringbauer_2022}.

\section{\texorpdfstring{$\mathcal{PT}$}{PT} symmetry}
\label{app:pt_invariance}

In this Appendix, we give a detailed proof of the $\mathcal{PT}$ symmetry, specializing to quantum ratchets composed of identical unitary gates~\eqref{eq:quantum_gate} with $V^T=V$. A particular example is the integrable ratchet, in which $V$ is the $R$-matrix~\eqref{eq:R_matrix}. 

To keep track of the ordering of spins, we first attach upper indices to the quantum gate~\eqref{eq:quantum_gate}, i.e., $U=U^{s_1,s_2}$. The space reflection $\ell\mapsto L-\ell+1$ acts as
\begin{align}
\label{eq:spin_exchange}
    \mathcal{P}(U^{s_1,s_2})=U^{s_2,s_1}.
\end{align}
The quantum gate~\eqref{eq:quantum_gate} exchanges the neighboring spins and can therefore be represented as
\begin{align}
\label{eq:decomposition}
    U^{s_1,s_2}=\sum_{\alpha,\beta=-s_1}^{s_1}\sum_{j,k=-s_2}^{s_2}U_{j\alpha,\beta k}\ket{j\alpha}\!\bra{\beta k},
\end{align} 
where the matrix elements $U_{j\alpha,\beta k}$ depend on $\tau\in\mathbb{R}$. Here and throughout the Appedix we use a convention in which Greek (Latin) indices enumerate the basis vectors of spin-$s_1$ (resp. spin-$s_2$) spaces, mainly to help distinguishing between the two different spins. 

A particular case of the gate~\eqref{eq:decomposition} is the permutation, $P^{s_1,s_2}$, whose matrix elements are $P_{j\alpha,\beta k}=\delta_{j,k}\delta_{\alpha,\beta}$, and one can verify that the following holds:
\begin{align}
\label{eq:act3}
    P^{s_1,s_2}U^{s_2,s_1}=U^{s_1,s_2}P^{s_2,s_1}.
\end{align}
Along with $(P^{s_1,s_2})^T=P^{s_2,s_1}=(P^{s_1,s_2})^{-1}$
and assuming the symmetry $V^T=V$ [cf. Eq.~\eqref{eq:symmetric_R} for the integrable case] we now have
\begin{align}
    \mathcal{P}(U^{s_1,s_2})&\!=\!P^{s_2,s_1}\!V^{s_2,s_1}\!=\!(V^{s_2,s_1}\! P^{s_1,s_2})^T\notag\\
    &\!=\!(P^{s_1,s_2}U^{s_2,s_1}\!P^{s_1,s_2})^T\!=\!(U^{s_1,s_2})^T,
\end{align}
i.e., Eq.~\eqref{eq:parity_action}.

In general, $U^T$ differs from $U$ and the circuit thus breaks the $\mathcal{P}$ symmetry. There however exists an additional antiunitary transformation $\mathcal{T}$, such that the full propagator given in Eqs.~\eqref{eq:propagator} and~\eqref{eq:two-steps} satisfies 
\begin{align}
    \mathcal{PT}(\mathbb{U})=\mathbb{U}^{-1}
\end{align}
and is hence $\mathcal{PT}$-symmetric.

To see this, we first note that the one-site lattice shift $\Pi_{s_1,s_2}$ in the backward direction, defined in Eq.~\eqref{eq:shift}, acts as 
\begin{align}
    \Pi_{s_1,s_2}&\ket{\alpha_1,j_1,\alpha_2,j_2,\ldots,\alpha_{L/2},j_{L/2}}=\notag\\
    &=\ket{j_1,\alpha_2,j_2,\alpha_3,\ldots,\alpha_{L/2},j_{L/2},\alpha_1}.
\end{align}
On the same lattice, composed of consecutive spins $s_1,s_2,s_1,\ldots,s_2$, the shift in the opposite direction is $\Pi^{-1}_{s_2,s_1}$, and it notably differs from $\Pi^{-1}_{s_1, s_2}$ when $s_1\ne s_2$ (it acts on a chain with a different ordering of spins). Since all of the unitary gates in the integrable ratchet are identical, we then have
\begin{align}
\label{eq:shifting_steps}
    \Pi^{-1}_{s_2, s_1}\mathbb{U}^{}_{\rm e}\Pi^{}_{s_1, s_2}=\mathbb{U}^{}_{\rm o},\quad\text{or}\quad \Pi^{}_{s_1, s_2}\mathbb{U}^{}_{\rm e}\Pi^{-1}_{s_2, s_1}=\mathbb{U}^{}_{\rm o},
\end{align}
where we have taken into account that the order in which the opposite lattice shifts are applied does not matter. Using this freedom of choice we can then write
\begin{align}
\label{eq:proof_PT}
    \underbrace{\mathbb{U}_{\rm o}^{-1}\mathbb{U}_{\rm e}^{-1}}_{\mathbb{U}^{-1}}
    &=(\Pi^{}_{s_2, s_1}\mathbb{U}_{\rm e}^{-1}\Pi_{s_1, s_2}^{-1})(\Pi^{}_{s_1, s_2}\mathbb{U}_{\rm o}^{-1}\Pi^{-1}_{s_2, s_1})\notag\\
    &=\Pi^{}_{s_2, s_1}K\mathbb{U}^T_{\rm e}\mathbb{U}^T_{\rm o}K\Pi^{-1}_{s_2, s_1}\notag\\
    &=\Pi^{}_{s_2, s_1}K\mathcal{P}(\,\underbrace{\mathbb{U}_{\rm e}\mathbb{U}_{\rm o}}_{\mathbb{U}}\,)K\Pi^{-1}_{s_2, s_1},
\end{align} 
where $K$ denotes the antiunitary conjugation. We have used unitarity of the evolution in passing to the second line, and Eq.~\eqref{eq:parity_action} in passing to the third line. Note that $\mathcal{P}$ acts simultaneously on all local unitary gates and, in general, cannot be written as an adjoint action of a product of some local operators.  Finally, defining the antiunitary time reversal $\mathcal{T}$ as the adjoint action of $\Pi_{s_2, s_1}K$, Eq.~\eqref{eq:proof_PT} demonstrates the $\mathcal{PT}$-symmetry of a brickwork ratchet composed of unitary gates $U=P^{s_1,s_2}V$ with $V^T=V$.

\section{Integrability}
\label{app:integrability}

Here, we show that the ratchet circuit composed of unitary gates~\eqref{eq:quantum_gate}, with $V=R^{s_1,s_2}(\tau)$ given in Eq.~\eqref{eq:R_matrix}, originates in the integrable family of transfer matrices~\eqref{eq:transfer_matrix}. Specifically, we will prove the transfer-matrix shift properties~\eqref{eq:odd_even_part} which, combined, yield the propagator $\mathbb{U}$ and the two-site lattice shift $\mathbb{T}$ in the backward (i.e., west) direction---see Eq.~\eqref{eq:propagator_from_transfer_matrices}. The two-site lattice shift explicitly reads
\begin{align}
    \mathbb{T}=T_{s_2}(\tfrac{\tau}{2})T_{s_1}(-\tfrac{\tau}{2})= \Pi_{s_2, s_1}\Pi_{s_1, s_2},
\end{align}
where Eqs.~\eqref{eq:odd_even_part} and~\eqref{eq:shifting_steps} have been used. It is formally an endomorphism, i.e., $\mathbb{T}\in{\rm End}[(\mathcal{H}_{s_1}\!\otimes\!\mathcal{H}_{s_2})^{\otimes L/2}]$,
and Eq.~\eqref{eq:propagator_from_transfer_matrices} implies the two-site shift invariance of the circuit, $[\,\mathbb{U},\mathbb{T}\,]=0$.
This follows from the commutation of transfer matrices for any pair of spins $s_1,s_2\in\mathbb{N}/2$, 
\begin{align}
    \left[\,T_{s_1}(\lambda),T_{s_2}(\mu)\,\right]=0,
\end{align}
which is a consequence of the Yang-Baxter equation~\eqref{eq:yang-baxter}.

The eigenvalues of $\mathbb{U}$ and $\mathbb{T}$ are, respectively, of the form $\exp(i 2\varepsilon_{\rm tot})$ and $\exp(-2i p_{\rm tot})$, where $\varepsilon_{\rm tot}$ is the total quasienergy and $p_{\rm tot}$ the total quasimomentum. After demonstrating the validity of the shift properties~\eqref{eq:odd_even_part} in Appendix~\ref{sec:shift-properties}, we will review the algebraic Bethe ansatz diagonalization of the transfer matrices $T_s(\lambda)$ in Appendix~\ref{sec:eigenvalues_propagator}. We will show that the total quasienergy and quasimomentum, resp. $\varepsilon_{\rm tot}$ and $p_{\rm tot}$, are extensive: they can be obtained as a sum of single-magnon contributions~\eqref{eq:quasienergy} and~\eqref{eq:quasimomentum}. Finally, explicit form of the Bethe equations will be reported in Appendix~\ref{sec:bethe_equations}.

\vspace{-1em}
\subsection{Propagator from transfer matrices}
\label{sec:shift-properties}

In this section we derive the transfer-matrix shift properties~\eqref{eq:odd_even_part} using the exchange relations
\begin{align}
    &P^{s_2,s_2}_{a,c}U^{s_2,s_1}_{b,c}=U^{s_1,s_2}_{a,b}P^{s_2,s_1}_{a,c},\label{eq:act1}\\
    &P^{s_1,s_2}_{a,c}U^{s_2,s_1}_{b,c}=U^{s_1,s_2}_{a,b}P^{s_1,s_1}_{a,c},\label{eq:act2}\\   
    &P^{s_1,s_2}_{a,c}P^{s_2,s_2}_{b,c}=P_{a,b}^{s_2,s_2}P_{a,c}^{s_1,s_2},\label{eq:act4}
\end{align}
and Eq.~\eqref{eq:act3}.
Note that these relations hold also when $U$ is substituted by its inverse $U^{-1}$, or by $P^{s_1,s_2}$ (recall that $U$ becomes a permutation when $\tau\to\infty$). The above relations are proven using a basis decomposition similar to the one in Eq.~\eqref{eq:decomposition}, with the first, second, and third index in the bra-ket notation corresponding to the spaces $a$, $b$, and $c$, respectively. For example, for Eq.~\eqref{eq:act1} we have
\begin{align}
    P^{s_2,s_2}_{a,c}U^{s_2,s_1}_{b,c}&=\sum_{\alpha,i,j}\ket{i \alpha j}\!\bra{ j \alpha i}\sum_{m,\gamma,\delta,k,l}U_{k\gamma ,\delta l}\ket{m \gamma k}\!\bra{m l\delta}\notag\\
    &=\sum_{\alpha,\delta,i,j,l}U_{i\alpha ,\delta l}\ket{i \alpha j}\!\bra{j l\delta}\notag\\
    &=\sum_{\alpha,\delta,i,j,l}U_{i\alpha ,\delta l}\ket{i \alpha j}\!\bra{\delta lj}P^{s_2,s_1}_{a,c}\notag\\
    &=U^{s_1,s_2}_{a,b}P^{s_2,s_1}_{a,c},
\end{align}
and similarly for other relations.

\subsubsection{North-west light-cone lattice shift} 

First, we derive the lattice shift in the north-west direction. To this end we consider the transfer matrix with a spin-$s_2$ auxiliary space, evaluated at $\lambda=\tau/2$. Using $R^{s_2,s_2}(0)=P^{s_2,s_2}$, $R^{s_1,s_2}(\tau)=P^{s_2,s_1}U^{s_1,s_2}$, and identification of the sites $0$ and $L$ due to periodic boundary conditions, we have
{\allowdisplaybreaks
\begin{widetext}
\begin{align}
    T_{s_2}(\tfrac{\tau}{2})&={\rm Tr}_{a}\prod_{1\le \ell\le L/2}^{\rightarrow}R^{s_1,s_2}_{2\ell-1,a}(\tau)R^{s_2,s_2}_{2\ell,a}(0)=
    {\rm Tr}_{a}\prod_{1\le \ell\le L/2}^{\rightarrow}P^{s_2,s_1}_{2\ell-1,a}U^{s_1,s_2}_{2\ell-1,a}P^{s_2,s_2}_{2\ell,a}=\notag\\
    &={\rm Tr}_{a}\prod_{1\le \ell\le L/2}^{\rightarrow}\underbrace{P^{s_2,s_2}_{2\ell-2,a}P^{s_2,s_1}_{2\ell-1,a}}_{\text{Eq.}~\eqref{eq:act1}}U^{s_1,s_2}_{2\ell-1,a}={\rm Tr}_{a}\prod_{1\le \ell\le L/2}^{\rightarrow}P^{s_1,s_2}_{2\ell-2,2\ell-1}\underbrace{P^{s_2,s_1}_{2\ell-2,a}U^{s_1,s_2}_{2\ell-1,a}}_{\text{Eq.}~\eqref{eq:act2}}=\notag\\
    &={\rm Tr}_{a}\prod_{1\le j\le L/2}^{\rightarrow}\underbrace{P^{s_1,s_2}_{2\ell-2,2\ell-1}U^{s_2,s_1}_{2\ell-2,2\ell-1}}_{\text{Eq.}~\eqref{eq:act3}}P^{s_2,s_2}_{2\ell-2,a}={\rm Tr}_{a}\prod_{1\le \ell\le L/2}^{\rightarrow}U^{s_1,s_2}_{2\ell-2,2\ell-1}P^{s_2,s_1}_{2\ell-2,2\ell-1}P^{s_2,s_2}_{2\ell-2,a}=\notag\\
    &=\mathbb{U}_{\rm e}\,{\rm Tr}_{a}\prod_{1\le \ell\le L/2}^{\rightarrow}P^{s_2,s_1}_{2\ell-2,2\ell-1}P^{s_2,s_2}_{2\ell-2,a}=\mathbb{U}_{\rm e}\Bigg(\prod_{1\le \ell\le L/2}^{\rightarrow}P^{s_2,s_1}_{2\ell-2,2\ell-1}\Bigg)\,{\rm Tr}_{a}\prod_{1\le \ell\le L/2}^{\rightarrow}P^{s_2,s_2}_{2\ell-2,a}=\notag\\
    &=\mathbb{U}_{\rm e}\Bigg(\prod_{1\le \ell\le L/2}^{\rightarrow}P^{s_2,s_1}_{2\ell-2,2\ell-1}\Bigg)\,\prod_{2\le \ell\le L/2}^{\rightarrow}P^{s_2,s_2}_{2\ell-2,L}=\mathbb{U}_{\rm e}\Bigg(\prod_{2\le \ell\le L/2}^{\rightarrow}P^{s_2,s_1}_{2\ell-2,2\ell-1}\Bigg)\underbrace{P^{s_2,s_1}_{L,1}\prod_{2\le \ell\le L/2}^{\rightarrow}P^{s_2,s_2}_{2\ell-2,L}}_{\text{Eq.}~\eqref{eq:act4}}=\notag\\
    &=\mathbb{U}_{\rm e}\Bigg(\prod_{2\le \ell\le L/2}^{\rightarrow}\underbrace{P^{s_2,s_1}_{2\ell-2,2\ell-1}P^{s_2,s_2}_{1,2\ell-2}}_{\text{Eq.}~\eqref{eq:act4}}\Bigg)P^{s_1,s_2}_{1,L}=\mathbb{U}_{\rm e}\Bigg(\prod_{2\le \ell\le L/2}^{\rightarrow}\underbrace{P^{s_2,s_2}_{1,2\ell-1}P^{s_1,s_2}_{2\ell-1,2\ell-2}}_{\text{Eq.}~\eqref{eq:act1}}\Bigg)P^{s_1,s_2}_{1,L}=\notag\\
    &=\mathbb{U}_{\rm e}\Bigg(\prod_{2\le \ell\le L/2}^{\rightarrow}P^{s_1,s_2}_{1,2\ell-2}P^{s_2,s_1}_{1,2\ell-1}\Bigg)P^{s_1,s_2}_{1,L}=\mathbb{U}_{\rm e}\,\Pi_{s_1, s_2},
\end{align}
\end{widetext}}
\noindent where $\Pi_{s_1, s_2}$ is a one-site lattice shift in the negative (i.e., west) direction. We have denoted which one of the exchange relations~\eqref{eq:act1},~\eqref{eq:act2},~\eqref{eq:act3}, and~\eqref{eq:act4} has to be used on a given pair of matrices in order to exchange them. On passing from the fourth to the fifth line we have also recognized that ${\rm Tr}^{}_a P^{s_2,s_2}_{L,a}=\mathbbm{1}$. This concludes the derivation of the north-west lattice shift reported on the left-hand side of Eq.~\eqref{eq:odd_even_part}.

\subsubsection{South-west light-cone lattice shift}

We now consider the south-west lattice shift: we will derive it from the transfer matrix with auxiliary spin $s_1$, evaluated at $\lambda=-\tau/2$. Equation~\eqref{eq:quantum_gate}, together with $V=R^{s_1,s_2}(\tau)$ and normalization~\eqref{eq:properties}, implies 
\begin{align}
    R^{s_2,s_1}(-\tau)&=[R^{s_2,s_1}(\tau)]^{-1}=[U^{s_2,s_1}]^{-1}P^{s_2,s_1}\notag\\
    &=[U^{-1}]^{s_1,s_2}P^{s_2,s_1},
\end{align}
where the fact that the quantum gate contains a permutation has been used in the last equality. With this in mind, we now have
{\allowdisplaybreaks
\begin{widetext}
\begin{align}
    T_{s_1}(-\tfrac{\tau}{2})&={\rm Tr}_{a}\prod_{1\le \ell\le L/2}^{\rightarrow}R^{s_1,s_1}_{2\ell-1,a}(0)R^{s_2,s_1}_{2\ell,a}(-\tau)={\rm Tr}_{a}\prod_{1\le \ell\le L/2}^{\rightarrow}\underbrace{P^{s_1,s_1}_{2\ell-1,a}[U^{-1}]^{s_1,s_2}_{2\ell,a}}_{\text{Eq.}~\eqref{eq:act1}}P^{s_2,s_1}_{2\ell,a}=\notag\\
    &={\rm Tr}_{a}\prod_{1\le \ell\le L/2}^{\rightarrow}[U^{-1}]^{s_2,s_1}_{2\ell-1,2\ell}P^{s_1,s_2}_{2\ell-1,a}P^{s_2,s_1}_{2\ell,a}=\mathbb{U}_{\rm o}^{-1}\, {\rm Tr}_{a}\prod_{1\le \ell\le L/2}^{\rightarrow}P^{s_1,s_2}_{2\ell-1,a}P^{s_2,s_1}_{2\ell,a}=\notag\\
    &=\mathbb{U}_{\rm o}^{-1}\, {\rm Tr}_{a}\underbrace{P_{1,a}^{s_1,s_2}\Bigg(\prod_{1\le \ell\le L/2-1}^{\rightarrow}P^{s_2,s_1}_{2\ell,a}P^{s_1,s_2}_{2\ell+1,a}\Bigg)}_{\text{Eqs.}~\eqref{eq:act2},~\eqref{eq:act1}}P_{L,a}^{s_2,s_1}=\notag\\
    &=\mathbb{U}_{\rm o}^{-1}\,\Bigg(\prod_{1\le \ell\le L/2-1}^{\rightarrow}P^{s_1,s_2}_{1,2\ell}P^{s_2,s_1}_{1,2\ell+1}\Bigg){\rm Tr}_{a}\underbrace{P^{s_1,s_2}_{1,a}P^{s_2,s_1}_{L,a}}_{\text{Eq.}~\eqref{eq:act2}}=\mathbb{U}_{\rm o}^{-1}\,\Bigg(\prod_{1\le \ell\le L/2-1}^{\rightarrow}P^{s_1,s_2}_{1,2\ell}P^{s_2,s_1}_{1,2\ell+1}\Bigg)P^{s_1,s_2}_{1,L}=\notag\\
    &=\mathbb{U}_{\rm o}^{-1}\,\Pi_{s_1, s_2}.
\end{align}
\end{widetext}}
\noindent In the fourth line we have again used ${\rm Tr}^{}_a P^{s_1,s_1}_{1,a}=\mathbbm{1}$.
This yields the lattice shift in the south-west direction, reported on the right-hand side of Eq.~\eqref{eq:odd_even_part}.

\subsection{Eigenvalues of transfer matrices and lattice shifts}
\label{sec:eigenvalues_propagator}

Here, we review the algebraic Bethe ansatz diagonalization of the integrable ratchet circuit. Following Refs.~\cite{babujian1983,babujian1982}, which discuss the higher-spin isotropic Heisenberg model, we write the transfer matrix~\eqref{eq:transfer_matrix} of the integrable quantum ratchet as
\begin{align}
\label{eq:transfer_matrix_and_monodromy}
    T_{s}(\lambda)={\rm Tr}_a M^{(s)}(\lambda)=\sum_{n=-s}^{s}M_{n,n}^{(s)}(\lambda),
\end{align}
where
\begin{align}
\label{eq:monodromy_matrix}
    M^{(s)}(\lambda)\equiv \prod_{1\le j\le L/2}^{\rightarrow}\!R^{s_1,s}_{2j-1,a}(\lambda\!+\!\tfrac{\tau}{2})R^{s_2,s}_{2j,a}(\lambda\!-\!\tfrac{\tau}{2})
\end{align}
is a $(2s+1)\times(2s+1)$ \emph{monodromy matrix} on the spin-$s$ auxiliary space labelled with $a$ (note that the auxiliary spin $s$ is not necessarily equal to any of the two physical spins $s_1,s_2$). Its entries $M_{m,n}^{(s)}(\lambda)$, with $m,n\in\{-s,-s+1,\ldots,s\}$, are operators acting on the full Hilbert space $\left(\mathcal{H}_{s_1}\otimes\mathcal{H}_{s_2}\right)^{\otimes L/2}$. 

The central role in diagonalizing the family of transfer matrices $T_s(\lambda)$ is played by the {\em fundamental transfer matrix}, associated with auxiliary spin $s=1/2$. It is obtained by tracing out the auxiliary space in the monodromy matrix
\begin{align}
\label{eq:shifted_fundamental_monodromy}
    M^{(1/2)}(\lambda-\tfrac{i}{2})=\begin{pmatrix}
        A(\lambda) & B(\lambda) \\
        C(\lambda) & D(\lambda)
    \end{pmatrix},
\end{align}
in which the spectral parameter has been shifted~\footnote{In what follows, shifting the spectral parameter allows us to reuse some of the results of the algebraic Bethe ansatz described in Ref.~\cite{faddeev1996}. There, the fundamental transfer matrix consists of {\em Lax operators}---the $R$-matrices with a shifted spectral parameter.}. In particular, an $N$-magnon eigenvector of $T_s(\lambda)$ reads (see, e.g., Ref.~\cite{faddeev1996}) 
\begin{align}
\label{eq:bethe_state}
    \ket{\{\lambda_j\}_{j=1}^N}=\prod_{j=1}^N B(\lambda_j)\ket{{\rm vac}},
\end{align}
where commuting operators $B(\lambda_j)$, with $j=1,\ldots, N$, have been applied to the vacuum (highest-weight) state
\begin{align}
\label{eq:vacuum}
    \ket{{\rm vac}}=\ket{\underbrace{s_1,s_2,s_1,s_2,\ldots,s_1,s_2}_{L}},
\end{align} 
and $\lambda_j$ denote the {\em rapidities} which satisfy nonlinear Bethe equations~\eqref{eq:bethe_equations}. 

Following Refs.~\cite{faddeev1996,babujian1983,babujian1982} one now requires two ingredients in order to obtain the eigenvalues of $T_s(\lambda)$:
\begin{enumerate}
    \item commutation relations that allow us to move $M_{n,n}^{(s)}(\lambda)$ past the sequence of operators $B(\lambda_j)$ in Eq.~\eqref{eq:bethe_state}, so that we can apply it to the vacuum state~\eqref{eq:vacuum};
    \item the vacuum eigenvalues of $M_{n,n}^{(s)}(\lambda)$.
\end{enumerate}
In the rest of Section~\ref{sec:eigenvalues_propagator} we first specify these two ingredients and then use them to obtain the eigenvalues of transfer matrices, of the propagator $\mathbb{U}$, and those of the lattice-shift operator $\mathbb{T}$. From them, we then determine the quasienergies and quasimomenta.

\vspace{-1em}
\subsubsection{Ingredients}

Firstly we describe the commutation relations between $M^{(s)}_{n,n}(\lambda)$ and $B(\mu)$. They are obtained by considering the appropriate matrix elements in the following relation,
\begin{align}
\label{eq:yb-algebra}
    M_a^{(1/2)}&(\mu\!-\!\tfrac{i}{2})M_b^{{(s)}}(\lambda)R_{a,b}^{1/2,s}(\lambda\!-\!\mu\!+\!\tfrac{i}{2})=\notag\\
    &=R_{a,b}^{1/2,s}(\lambda\!-\!\mu\!+\!\tfrac{i}{2})M_b^{{(s)}}(\lambda)M_a^{(1/2)}(\mu\!-\!\tfrac{i}{2}),
\end{align}
which holds by virtue of the Yang-Baxter equation~\eqref{eq:yang-baxter}. Here, we have temporarily attached lower indices $a$ and $b$ to monodromy matrices. They denote the auxiliary spaces of respective spins $1/2$ and $s$. Equating the matrix elements on both sides of Eq.~\eqref{eq:yb-algebra} yields relations between the entries of the two monodromy matrices. The coefficients in these relations are the elements of the $R$-matrix $R_{a,b}^{1/2,s}(\lambda-\mu+i/2)$. Crucially, they do not depend on the inhomogeneity parameter $\tau$ in the monodromy matrix~\eqref{eq:monodromy_matrix}.

The relation we are interested in involves $M^{(s)}_{n,n}(\lambda)$ and $B(\mu)$. It was reported in Refs.~\cite{babujian1982,babujian1983} and reads
{\allowdisplaybreaks
\begin{widetext}
\begin{align}
\label{eq:babujian}
    M^{(s)}_{n,n}(\lambda)B(\mu)=&\,c^{(s)}_{0}(\lambda-\mu+\tfrac{i}{2};n)B(\mu)M^{(s)}_{n,n}(\lambda)+c^{(s)}_{1}(\lambda-\mu+\tfrac{i}{2};n)M^{(s)}_{n,n-1}(\lambda)A(\mu)+\notag\\[1em]
    &\,+c^{(s)}_{2}(\lambda-\mu+\tfrac{i}{2};n)M^{(s)}_{n+1,n}(\lambda)D(\mu)+c^{(s)}_{3}(\lambda-\mu+\tfrac{i}{2};n)M^{(s)}_{n+1,n-1}(\lambda)C(\mu),
\end{align}
\end{widetext}}
\noindent where the $\tau$-independent coefficients $c^{(s)}_j(\lambda;n)$ are~\footnote{As a result of a different choice of spectral parameter in the $R$-matrix, the coefficients reported in Eq.~(37) in Ref.~\cite{babujian1983} are reproduced by substituting our spectral parameter $\lambda$ with $\lambda/i$. Moreover, the first coefficient (i.e., $c^{(s)}_0$) in Ref.~\cite{babujian1983} seems to contain a typo: the correct coefficient, coinciding with ours up to an imaginary unit in the spectral parameter, is instead reported in Eq.~(20) in Ref.~\cite{babujian1982}.}
\begin{align}
\begin{aligned}
    &c^{(s)}_{0}(\lambda;n)=\frac{(\lambda-i s-\tfrac{i}{2})(\lambda+i s+\tfrac{i}{2})}{(\lambda+i n-\tfrac{i}{2})(\lambda+i n +\tfrac{i}{2})},\\[1em]
    &c^{(s)}_{1}(\lambda;n)=\frac{\sqrt{(n+s)(n-s-1)}}{\lambda+i n -\tfrac{i}{2}},\\[1em]
    &c^{(s)}_{2}(\lambda;n)=-\frac{\sqrt{(n-s)(n+s+1)}}{\lambda+i n+\tfrac{i}{2}},\\[1em]
    &c^{(s)}_{3}(\lambda;n)=-\frac{\sqrt{(s^2-n^2)[(s+1)^2-n^2]}}{(\lambda+i n+\tfrac{i}{2})(\lambda+i n-\tfrac{i}{2})}.
\end{aligned}
\end{align}
Note that the spectral parameter $\mu$ in the functions $c^{(s)}_{j}$ in Eq.~\eqref{eq:babujian} has been shifted as $\mu\mapsto\mu-i/2$. This is because operators $A(\mu)$, $B(\mu)$, $C(\mu)$, and $D(\mu)$ are inferred from the monodromy matrix~\eqref{eq:shifted_fundamental_monodromy} with the same shift in the spectral parameter. Of paricular importance is the first term on the right-hand side of Eq.~\eqref{eq:babujian}: it exchanges $M_{n,n}^{(s)}(\lambda)$ with a magnon creation operator $B(\mu)$, producing a factor given by the function $c^{(s)}_0(\lambda-\mu+i/2;n)$. The latter will ``dress'' the vacuum eigenvalue of $M^{(s)}_{n,n}(\lambda)$.

We now consider the vacuum eigenvalues of the monodromy's diagonal entries $M^{(s)}_{n,n}(\lambda)$. For our purposes, it will suffice to consider  $s\in\{s_1,s_2\}$, but the discussion below remains valid for other values of the auxiliary spin $s$. Following Ref.~\cite{faddeev1996}, the vacuum eigenvalues are obtained by noting that the $R$-matrix becomes upper-triangular when applied to the vacuum state $\ket{{\rm vac}}$. Specifically, denoting the physical spin by $s_p$ ($s_p=s_1$ for odd-site indices $j$ and $s_p=s_2$ for even-site indices $j$), we have~\footnote{In Eq.~\eqref{eq:R-matrix_diagonal} we have exchanged the order of the spaces labeled by $j$ and $a$. In the new ordering we can use the standard convention for the matrix representation of the tensor product, in which the elements of the matrix act on the space with the second index, i.e., $j$.}
\begin{widetext}
    \begin{align}
    \label{eq:R-matrix_diagonal}
    R^{s,s_p}_{a,j}(\lambda)\ket{s_p}_j=
    \left(
    \begin{array}{c c c c c}
    \alpha^{(s)}_{s}(\lambda;s_p)\ket{s_p}_j 
    &  
    &  
    &  
    & 
    \\
    & 
    \alpha^{(s)}_{s-1}(\lambda;s_p)\ket{s_p}_j 
    &  
    & 
    \bigasterisk
    &  
    \\
    &  
    &
    \ddots
    &  
    &  
    \\
    &  
    \bigzero
    &  
    & 
    \alpha^{(s)}_{-s+1}(\lambda;s_p)\ket{s_p}_j
    & 
    \\
    & 
    &  
    & 
    & 
    \alpha^{(s)}_{-s}(\lambda;s_p)\ket{s_p}_j
    \end{array}\right).
\end{align}
\end{widetext}
Here, $\alpha^{(s)}_n(\lambda;s_p)$, with $n\in\{-s,-s+1,\ldots,s\}$, are some functions which can be computed explicitly and satisfy the following properties:
\begin{enumerate}
    \item $\alpha^{(s)}_n(0;s)=0$ if $n\ne s$;
    \item for any $s_p$, $s$, and $\lambda$, we have $\alpha^{(s)}_{s}(\lambda;s_p)=1$.
\end{enumerate}

When acting on the vacuum state~\eqref{eq:vacuum}, the monodromy $M^{(s)}(\lambda)$ becomes a product of $R$-matrices of the form~\eqref{eq:R-matrix_diagonal}. Since the latter are upper-triangular, the diagonal entries of $M^{(s)}(\lambda)$ satisfy
\begin{align}
\label{eq:monodromy_diagonal}
    M^{(s)}_{n,n}\!(\lambda)\!\ket{{\rm vac}}\!=\!\left[\alpha_n^{(s)}\!(\lambda_+;s_1)\alpha_n^{(s)}\!(\lambda_-;s_2)\right]^{\tfrac{L}{2}}
    \!\ket{{\rm vac}},
\end{align}
where we have used a shorthand notation $\lambda_\pm=\lambda\pm\tau/2$.

\subsubsection{Eigenvalues of the transfer matrix}

Applying the monodromy $M^{(s)}(\lambda)$ to a Bethe state~\eqref{eq:bethe_state},
its diagonal entries~\eqref{eq:monodromy_diagonal} get ``dressed'' due to commutation relations~\eqref{eq:babujian}. Tracing the monodromy over the auxiliary space, as per Eq.~\eqref{eq:transfer_matrix_and_monodromy}, we then obtain
\begin{align}
    T_{s}(\lambda)\ket{\{\lambda_j\}}=&\,\Lambda_{s}(\lambda;\{\lambda_j\})\ket{\{\lambda_j\}}\notag\\[1em]
    &+\text{unwanted terms},\label{eq:tm_eigenvalue-line1}\\[1em]
    \Lambda_{s}(\lambda;\{\lambda_j\})=&\sum_{n=-s}^{s}[\alpha_n^{(s)}(\lambda_+;s_1)\alpha_n^{(s)}(\lambda_-;s_2)]^{\tfrac{L}{2}}\notag\\
    &\times\prod_{j=1}^N c^{(s)}_0(\lambda-\lambda_j+\tfrac{i}{2};n),\label{eq:tm_eigenvalue-line2}
\end{align}
where $s=s_1,s_2$. Assuming that rapidities $\lambda_j$ satisfy Bethe equations~\eqref{eq:bethe_equations}, the ``unwanted terms'' in Eq.~\eqref{eq:tm_eigenvalue-line1} disappear---see, e.g., Refs.~\cite{faddeev1996,babujian1982}. $\Lambda_{s}(\lambda;\{\lambda_j\})$ given in Eq.~\eqref{eq:tm_eigenvalue-line2} are then the eigenvalues of the transfer matrix with a spin-$s$ auxiliary space.

\subsubsection{Eigenvalues of the propagator and the lattice shift}

The propagator and the lattice-shift operator can be written in terms of transfer matrices---cf. Eq.~\eqref{eq:propagator_from_transfer_matrices}. We exploit this to obtain
\begin{align}
    \mathbb{U}\ket{\{\lambda_j\}}&=\frac{\Lambda_{s_2}(\tfrac{\tau}{2};\{\lambda_j\})}{\Lambda_{s_1}(-\tfrac{\tau}{2};\{\lambda_j\})}\ket{\{\lambda_j\}}=e^{i2 \varepsilon_{\rm tot}}\ket{\{\lambda_j\}}.
\end{align}
Since $\alpha^{(s)}_{n\ne s}(0;s)=0$ and $\alpha^{(s)}_{s}(\lambda;s_p)=1$, for any $s_p$, many terms in the eigenvalue~\eqref{eq:tm_eigenvalue-line2} disappear, leaving us with
\begin{align}
    e^{i2 \varepsilon_{\rm tot}}&=\prod_{j=1}^N \frac{c^{(s_2)}_0(\tfrac{\tau}{2}\!-\!\lambda_j\!+\!\tfrac{i}{2};s_2)}{c^{(s_1)}_0(\!-\!\tfrac{\tau}{2}\!-\!\lambda_j\!+\!\tfrac{i}{2};s_1)}\notag\\
    &=\prod_{j=1}^N\frac{(\lambda_j\!+\!\tfrac{\tau}{2}\!-\!i s_1)(\lambda_j\!-\!\tfrac{\tau}{2}\!+\!i s_2)}{(\lambda_j\!+\!\tfrac{\tau}{2}\!+\!i s_1)(\lambda_j\!-\!\tfrac{\tau}{2}\!-\!i s_2)}.
\end{align}
The quasienergy is extensive, i.e., $\varepsilon_{\rm tot}=\sum_{j=1}^N \varepsilon(\lambda_j)$, and the single-magnon quasienergies~\eqref{eq:quasienergy} are obtained from
\begin{align}
\label{eq:app_quasienergy}
e^{i2\varepsilon(\lambda)}=
\frac{(\lambda_+ -i s_1)(\lambda_- +i s_2)}{(\lambda_+ +i s_1)(\lambda_- -i s_2)}.
\end{align}

The eigenvalue of the two-site lattice shift operator is obtained similarly [see Eq.~\eqref{eq:propagator_from_transfer_matrices}]:
\begin{align}
    \mathbb{T}\ket{\{\lambda_j\}}&=\Lambda_{s_2}(\tfrac{\tau}{2};\{\lambda_j\})\Lambda_{s_1}(-\tfrac{\tau}{2};\{\lambda_j\})\ket{\{\lambda_j\}}\notag\\[1em]
    &=e^{-i2 p_{\rm tot}}\ket{\{\lambda_j\}}.
\end{align}
It reads
\begin{align}
    e^{-i2 p_{\rm tot}}&=\prod_{j=1}^N c^{(s_2)}_0(\tfrac{\tau}{2}\!-\!\lambda_j\!+\!\tfrac{i}{2};s_2)c^{(s_1)}_0(\!-\!\tfrac{\tau}{2}\!-\!\lambda_j\!+\!\tfrac{i}{2};s_1)\notag\\
    &=\prod_{j=1}^N\frac{(\lambda_j+\tfrac{\tau}{2}+i s_1)(\lambda_j-\tfrac{\tau}{2}+i s_2)}{(\lambda_j+\tfrac{\tau}{2}-i s_1)(\lambda_j-\tfrac{\tau}{2}-i s_2)},
\end{align}
and the total quasimomentum is extensive as well: $p_{\rm tot}=\sum_{j=1}^N p(\lambda_j)$. We can identify the single-magnon quasimomentum~\eqref{eq:quasimomentum} from
\begin{align}
\label{eq:app_quasimomentum}
    e^{-2i p(\lambda)}=\frac{(\lambda_+ +i s_1)(\lambda_- +i s_2)}{(\lambda_+ -i s_1)(\lambda_- -i s_2)}.
\end{align}

\subsection{Bethe equations}
\label{sec:bethe_equations}

The entries of the fundamental monodromy matrix~\eqref{eq:shifted_fundamental_monodromy} satisfy commutation relations~\eqref{eq:babujian} specialized to the case $\lambda\mapsto\lambda-i/2$ and $s=1/2$:
\begin{align}
\begin{aligned}
    A(\lambda)B(\mu)=&c^{(1/2)}_0(\lambda-\mu;\tfrac{1}{2})B(\mu)A(\lambda)\\
    &+c^{(1/2)}_1(\lambda-\mu;\tfrac{1}{2})B(\lambda)A(\mu),\\
    D(\lambda)B(\mu)=&c^{(1/2)}_0(\lambda-\mu;-\tfrac{1}{2})B(\mu)D(\lambda)\\
    &+c^{(1/2)}_2(\lambda-\mu;-\tfrac{1}{2})B(\lambda)D(\mu).\\
\end{aligned}
\end{align}
Following Ref.~\cite{faddeev1996}, these relations allow us to identify the two-magnon scattering amplitude as 
\begin{align}
\label{eq:scattering_matrix}
    \mathcal{S}(\lambda-\mu)=\frac{c_0^{(1/2)}(\lambda-\mu;-\tfrac{1}{2})}{c_0^{(1/2)}(\lambda-\mu;\tfrac{1}{2})}=\frac{\lambda-\mu+i}{\lambda-\mu-i}.
\end{align}
Using Eqs.~\eqref{eq:app_quasimomentum} and~\eqref{eq:scattering_matrix} in the quantization condition~\eqref{eq:bethe_equations} for the single-magnon quasimomenta, we rewrite the Bethe equations in an explicit form as
\begin{align}
\label{eq:app_BAE}
    \left[\frac{(\lambda_j\!+\!\tfrac{\tau}{2}\!+\!i s_1)(\lambda_j\!-\!\tfrac{\tau}{2}\!+\!i s_2)}{(\lambda_j\!+\!\tfrac{\tau}{2}\!-\!i s_1)(\lambda_j\!-\!\tfrac{\tau}{2}\!-\!i s_2)}\right]^{\tfrac{L}{2}}=\prod^N_{\substack{k=1\\k\ne j}}\frac{\lambda_j\!-\!\lambda_k\!+\!i}{\lambda_j\!-\!\lambda_k\!-\!i}.
\end{align}

\section{Thermodynamic Bethe ansatz}
\label{app:TBA}

This appendix describes the Bethe ansatz and its solution in the thermodynamic (TD) limit $N,L\to\infty$, at a fixed ratio $N/L$, where $N$ is the particle number and $L$ the system size. In Appendix~\ref{app:BY-equations} we describe the Bethe ansatz equations in the thermodynamic limit (the so-called {\em Bethe-Yang equation}). In Appendix~\ref{app:TBA-dressing} we define the dressing operation, i.e., a renormalization of the bare charge carried by the quasiparticles, necessitated by their interaction. Lastly, Appendix~\ref{app:TBA-solution} describes a novel solution of the Bethe-Yang equation with a particular emphasis on the computation of the drift velocity.

\subsection{Bethe-Yang equation}
\label{app:BY-equations}

For large $L$, the solutions of Bethe ansatz equations organize into the so-called ``$m$-strings'', i.e., complexes of $m$ bound magnons carrying $m$ quanta of magnetization, referred to as the bare charge $q_m=m$. Let there be $M_m$ $m$-strings in a particular solution of the Bethe ansatz equations. Up to corrections exponentially small in $L$, the corresponding rapidities read
\begin{align}
\label{eq:strings}
    \lambda^{m,j}_\alpha=\lambda^m_\alpha+\tfrac{i}{2}(m+1-2j),\quad j=1,\ldots,m,
\end{align}
where $\lambda^m_\alpha\in\mathbb{R}$, for $\alpha=1,\ldots, M_m$, are the {\em string centers} which become densely distributed in the TD limit. Their distributions satisfy an appropriate TD limit of the Bethe ansatz equations, which we will refer to as the {\em Bethe-Yang equation}.

In a chain of alternating spins, the quasienergies~\eqref{eq:quasienergy} and quasimomenta~\eqref{eq:quasimomentum} are sums of contributions from sublattices of spins $s_1$ and $s_2$. As will become clear later on, the same holds for the distribution of the string centers, which is why we will first consider the Bethe-Yang equation of a homogeneous spin-$s$ Heisenberg chain. The latter's Bethe equations  are obtained by setting $s_1=s_2=s$ and $\tau=0$ in Eq.~\eqref{eq:app_BAE}. Inserting the string form~\eqref{eq:strings} and multiplying the equations for $j=1,\ldots,m$, one obtains the equations for the string centres $\lambda^m_\alpha$ (see Ref.~\cite{babujian1983} and also Section 8.2 in Ref.~\cite{takahashi1999} for an analogous procedure in a spin-$1/2$ Heisenberg chain). Taking then the logarithm, one obtains the Bethe-Yang equation in the TD limit.

Specifically, let $b\equiv2s$ and let $\rho_m^{{\rm tot}(b)}(\lambda)$ be the total state densities, defined so that $L\rho_m^{{\rm tot}(b)}(\lambda){\rm d}\lambda$ is the number of available $m$-string centers in the {\em rapidity} interval $[\lambda,\lambda+{\rm d}\lambda)\subset\mathbb{R}$. To define a macrostate of the system, one in addition requires the distribution of the {\em occupied} $m$-string centers, $\rho^{(b)}_m(\lambda)$, or equivalently, the occupancy ratio $n_m(\lambda)\equiv\rho^{(b)}_m(\lambda)/\rho_{m}^{{\rm tot}(b)}(\lambda)$. Treating $m\in \mathbb{N}$ as a discrete index and $\lambda\in\mathbb{R}$ as a continuous one, we can define a matrix $\mathbf{n}$ with elements $n_m(\lambda)\delta_{m,m'}\delta(\lambda-\lambda')$, and a vector $\boldsymbol{\rho}^{{\rm tot}(b)}$ with components $\rho_m^{{\rm tot}(b)}(\lambda)$. Bethe-Yang equation reads
\begin{align}
\label{eq:app_bethe-yang}
(\mathbf{1}+\mathbf{K}\mathbf{n})\boldsymbol{\rho}^{{\rm tot}(b)}=\boldsymbol{K}^{(b)},
\end{align}
where the components of the vector $\boldsymbol{K}^{(b)}$ are the strings' bare quasimomentum derivatives
\begin{align}
     K_m^{(b)}(\lambda)\!\equiv\!\frac{1}{2\pi}\partial_\lambda p_m^{(b)}(\lambda)\!=\!\sum_{\ell=1}^{\min(m,b)}\!K_{|m-b|-1+2\ell}(\lambda),
\end{align}
expressed in terms of functions
\begin{align}
    K_{m\ge 1}(\lambda)=\frac{1}{2\pi} \frac{m}{\lambda^2+(m/2)^2},\qquad K_0(\lambda)\equiv 0.
\end{align}
Finally, $\mathbf{K}$ is a matrix with elements
\begin{align}
\label{eq:app_integral_kernel}
    \operatorname{K}_{m,\ell}(\lambda\!-\!\lambda')&=\sum_{j=|m-\ell|/2}^{(m+\ell)/2-1}[K_{2j}(\lambda\!-\!\lambda')+K_{2j+2}(
    \lambda\!-\!\lambda')]=\notag\\
    &=K_{m-1}^{(\ell)}(\lambda\!-\!\lambda')+K_{m+1}^{(\ell)}(\lambda\!-\!\lambda').
\end{align}
In this compact notation the matrix product involves summation over the discrete mode index $m$ and integration over the continuous rapidity $\lambda$ (i.e., convolution). 

To determine both $\boldsymbol{\rho}^{{\rm tot}(b)}$ and $\mathbf{n}$, we require another set of equations, which are obtained by maximizing the thermodynamic free energy. A crucial simplification occurs in the thermodynamic state described by the density matrix~\eqref{eq:Gibbs-state}. There, $\mathbf{n}$ is determined from thermodynamic equations that do not depend on spin $s$, and we can therefore use $\mathbf{n}$ computed in a homogeneous spin-$1/2$ Heisenberg model. Moreover, in such a state the occupancy ratio does not depend on the rapidity either---it reads
\begin{align}
\label{eq:app_occupancy}
    n_m=\frac{1}{\mathcal{X}^2_m(\mu)},
\end{align}
where
\begin{align}
\label{eq:app_character}
    \mathcal{X}_m(\mu)=\frac{\sinh\left([m+1]\tfrac{\mu}{2}\right)}{\sinh\left(\tfrac{\mu}{2}\right)}
\end{align}
is an $\mathfrak{su}(2)$ character (see, e.g., Section 8.4 in Ref.~\cite{takahashi1999} or the Supplemental material of Ref.~\cite{ilievski2018}). 

Crucially, the Bethe-Yang equation~\eqref{eq:app_bethe-yang} is linear in  $\boldsymbol{\rho}^{{\rm tot}(b)}$, and  $\mathbf{n}$ does not depend on the spin. Hence, since the bare quasimomentum  in the integrable ratchet is a sum of two contributions, $[K_m^{(b_1)}(\lambda_+)+K_m^{(b_2)}(\lambda_-)]/2$, the solution of Eq.~\eqref{eq:app_bethe-yang} will be a sum as well: $[\rho_m^{{\rm tot}(b_1)}(\lambda_+)+\rho_m^{{\rm tot}(b_2)}(\lambda_-)]/2$.

\subsection{Dressing and screening operations} 
\label{app:TBA-dressing}

Bethe-Yang equation~\eqref{eq:app_bethe-yang} is a specific example of the dressing equation~\cite{alvaredo2016}
\begin{align}
\label{eq:app_dressing}
    (\mathbf{1}+\mathbf{K}\mathbf{n})\boldsymbol{q}^{\rm dr}=\boldsymbol{q}\quad \Rightarrow \quad \boldsymbol{q}^{\rm dr}= (\mathbf{1}+\mathbf{K}\mathbf{n})^{-1}\boldsymbol{q},
\end{align}
which encodes the effect of the interactions between the quasiparticles on the charge carried by them. In particular, according to Eq.~\eqref{eq:app_bethe-yang} the total state density is associated with a dressed derivative of the quasimomentum:
\begin{align}
\label{eq:app_dressed-momentum}
    2\pi \boldsymbol{\rho}^{{\rm tot}(b)}\equiv(\boldsymbol{p}^{(b)}{'})^{\rm dr}.
\end{align}

\setcounter{definition}{0}
\begin{definition}
The dressing equation for the quasiparticle magnetization $q_m=m$ is solved by 
\begin{align}
    q_m^{\rm dr}=\partial_\mu \log(n_m^{-1}-1),   
\end{align}
where $n_m$ is given in Eq.~\eqref{eq:app_occupancy}.
\end{definition}

\begin{definition}
According to Eq.~\eqref{eq:app_integral_kernel} the kernel $\operatorname{K}_{m,\ell}$ is a sum of bare quasimomenta and can therefore be dressed as well. We have $\mathbf{K}^{\rm dr}=(\mathbf{1}+\mathbf{K}\mathbf{n})^{-1}\mathbf{K}$ or, element-wise, 
\begin{align}
\label{eq:app_dressed-kernel}
    \operatorname{K}^{\rm dr}_{m,\ell}(\lambda)=\rho_{m}^{{\rm tot}(\ell-1)}(\lambda)+\rho_{m}^{{\rm tot}(\ell+1)}(\lambda),
\end{align} 
where $\rho_0^{{\rm tot}(b)}(\lambda)\equiv 0$ and $\rho_m^{{\rm tot}(0)}(\lambda)\equiv 0$. The derivation of Eq.~\eqref{eq:app_dressed-kernel} is outlined in the next section. We now note that, since $\mathbf{K}^{\rm dr}\mathbf{n}$ is an operator series in $\mathbf{K}\mathbf{n}$, it commutes with $\mathbf{1}+\mathbf{K}\mathbf{n}$, and the dressing operator is then simply found to be $(\mathbf{1}+\mathbf{K}\mathbf{n})^{-1}=\mathbf{1}-\mathbf{K}^{\rm dr}\mathbf{n}$. It then follows that
\begin{align}
    \boldsymbol{q}^{\rm scr}\equiv\boldsymbol{q}-\boldsymbol{q}^{\rm dr}=\mathbf{K}^{\rm dr}\mathbf{n}\,\boldsymbol{q},
\end{align}
which may be understood as a {\em screened charge}, and which we use in the expression for the third scaled cumulant of the time-integrated spin current in Section~\ref{sec:fluctuations}.
Written out explicitly in terms of the convolution, the screened charge reads
\begin{align}
    q^{\rm scr}_m(\lambda)=\sum_{\ell=1}^\infty\int{\rm d}\lambda' \operatorname{K}^{\rm dr}_{m,\ell}(\lambda-\lambda')n_\ell(\lambda')q_\ell(\lambda').
\end{align}
\end{definition}

\subsection{Solution of the Bethe-Yang equation}
\label{app:TBA-solution}

Bethe-Yang equation~\eqref{eq:app_bethe-yang} can be rewritten as a three-point recurrence relation for the total state densities $\rho^{{\rm tot}(b)}_m(\lambda)$. Its derivation is based on the observation that the Fourier-transformed kernels $\hat{K}_m(k)\equiv\int{\rm d}\lambda e^{-ik\lambda}K_m(\lambda)=e^{-m|k|/2}$ satisfy 
\begin{align}
    \hat{K}_m-\hat{\mathfrak{s}}(\hat{K}_{m+1}+\hat{K}_{m-1})=\delta_{m,1} \hat{\mathfrak{s}},
\end{align}
where $\hat{\mathfrak{s}}(k)=[2\cosh(k/2)]^{-1}$ and $\hat{K}_0\equiv 0$~\cite{ilievski2018}. In the rapidity space one then has
\begin{align}
\label{eq:app_recurrence1}
    \sum_m(\delta^{}_{n,m}\delta-I^{}_{n,m}\mathfrak{s})\star K^{(b)}_m=\delta^{}_{n,b}\mathfrak{s},  
\end{align} 
where $(f\star g)(\lambda)=\int{\rm d}\lambda' f(\lambda-\lambda')g(\lambda')$ denotes the convolution, $\delta$ is the delta-function, and $I_{n,m}=\delta_{n,m+1}+\delta_{n,m-1}$. 

Expressing the integral kernel~\eqref{eq:app_integral_kernel} as $\operatorname{K}_{m,\ell}\!=\!\sum_n I_{m,n}K_n^{(\ell)}$ in the Bethe-Yang equation~\eqref{eq:app_bethe-yang}, applying on it the matrix with elements $\delta_{n,m}\delta(\lambda\!-\!\lambda')\!-\!I_{n,m}\mathfrak{s}(\lambda\!-\!\lambda')$, and invoking Eq.~\eqref{eq:app_recurrence1}, we then obtain the  recurrence for the total state density. In the Fourier space it reads
\begin{align}
\label{eq:app_recurrence2}
    \hat{\mathfrak{s}}^{-1} \hat{\rho}_m^{{\rm tot}(b)}-\sum_{\ell=1}^{\infty}I_{m,\ell}(1-n_\ell)\hat{\rho}_\ell^{{\rm tot}(b)}=\delta_{m,b},
\end{align}
where we have used that $n_m$ in the state described by the density matrix~\eqref{eq:Gibbs-state} does not depend on the rapidity $\lambda$.
We note that the recurrence satisfied by the dressed kernel in Eq.~\eqref{eq:app_dressed-kernel} can be obtained similarly. It reads
\begin{align}
\label{eq:kernel-dressing}
    \hat{\mathfrak{s}}^{-1} \hat{\operatorname{K}}^{\rm dr}_{m,j}-\sum_{\ell=1}^{\infty}I_{m,\ell}(1-n_\ell)\hat{\operatorname{K}}^{\rm dr}_{\ell,j}=I_{m,j},
\end{align}
which is the same as Eq.~\eqref{eq:app_recurrence2} multiplied by $I_{b,j}$ from the left, if we identify $\hat{\operatorname{K}}^{\rm dr}_{m,j}=\sum_b\hat{\rho}_m^{{\rm tot}(b)}I_{b,j}$. The latter is equivalent to Eq.~\eqref{eq:app_dressed-kernel}.

In the following we describe how to obtain a closed-form solution of the recurrence~\eqref{eq:app_recurrence2}.

\subsubsection{Total state density}

Let us introduce $z\equiv e^{\mu/2}$ and consider only $k>0$, since Eq.~\eqref{eq:app_recurrence2} is invariant under the transformation $k\mapsto -k$. The occupancy ratio is given in Eq.~\eqref{eq:app_occupancy}, and it only depends on $z$ through the $\mathfrak{su}(2)$ character~\eqref{eq:app_character}. Setting the source term on the right-hand side of Eq.~\eqref{eq:app_recurrence2} to zero, we first find two homogeneous solutions
\begin{align}
    \phi^{(b)}_{m;\pm}\!(k;\!z)\!=\!\frac{\mathcal{X}_{m}(z)}{\mathcal{X}_{1}(z)}
\!\left(\frac{e^{\pm(m-1)k/2}}{\mathcal{X}_{m-1}(z)}\!-\!\frac{e^{\pm(m+1)k/2}}{\mathcal{X}_{m+1}(z)}\right).
\end{align}
With the source term present, homogeneous solutions should be glued together at the index $m=b$. In particular, since the total state density should obey $\lim_{|\lambda|\to\infty}\rho^{{\rm tot}(b)}_m(\lambda)=0$, we will take the following ansatz:
\begin{align}
\label{eq:app_ansatz}
    \hat{\rho}^{{\rm tot}(b)}_m=\begin{cases}
        \mathcal{A}^{(b)}\phi^{(b)}_{m;-}+\mathcal{B}^{(b)}\phi^{(b)}_{m;+} & m<b, \\
        \mathcal{C}^{(b)}\phi^{(b)}_{m;-} & m\ge b,
    \end{cases}
\end{align}
where the coefficients $\mathcal{A}^{(b)}$, $\mathcal{B}^{(b)}$, and $\mathcal{C}^{(b)}$ are some functions of $k$ and $z$, determined by plugging the ansatz into the recurrence~\eqref{eq:app_recurrence2}. In particular, two equations at the index value $m=b$ and one at $m=1$ suffice to determine the three coefficients. The relevant equations~\eqref{eq:app_recurrence2} are
\begin{align}
    \begin{aligned}
        2\cosh(\tfrac{k}{2})\hat{\rho}^{{\rm tot}(b)}_1\!-\!\frac{\mathcal{X}_2^2\!-\!1}{\mathcal{X}_2^2}\hat{\rho}^{{\rm tot}(b)}_2\!&=\!0,\\
        2\cosh(\tfrac{k}{2})\hat{\rho}^{{\rm tot}(b)}_{b-1}\!-\!\frac{\mathcal{X}_b^2\!-\!1}{\mathcal{X}_b^2}\hat{\rho}^{{\rm tot}(b)}_b\!-\!\frac{\mathcal{X}_{b-2}^2\!-\!1}{\mathcal{X}_{b-2}^2}\hat{\rho}^{{\rm tot}(b)}_{b-2}\!&=\!0,\\
        2\cosh(\tfrac{k}{2})\hat{\rho}^{{\rm tot}(b)}_{b}\!-\!\frac{\mathcal{X}_{b+1}^2\!-\!1}{\mathcal{X}_{b+1}^2}\hat{\rho}^{{\rm tot}(b)}_{b+1}\!-\!\frac{\mathcal{X}_{b-1}^2\!-\!1}{\mathcal{X}_{b-1}^2}\hat{\rho}^{{\rm tot}(b)}_{b-1}\!&=\!1,
    \end{aligned}
\end{align}
and the coefficients that solve them take the following form:
\begin{widetext}
\begin{align}
    \begin{aligned}
    \label{eq:app_coefs}
    \mathcal{A}^{(b)}(k\!;z) &= -\frac{(1+z^{2})(z^{2}+z^{2(b+1)}[e^{k}z^{2}-1]-e^{k})}{(e^{k}-1)(z^{2(b+1)}-1)(z^{4}-z^{2}[e^{k}+e^{-k}]+1)}e^{-(b+2)k/2},\qquad
    \mathcal{B}^{(b)}(k\!;z)= -e^{k}\mathcal{A}^{(b)}(k;z),\\
    \mathcal{C}^{(b)}(k\!;z) &= \frac{(1+z^{2})\big\{z^{2(b+1)}[e^{(b+2)k}-z^{2}e^{k}(e^{bk}-1)-1]+z^{2}(1-e^{(b+2)k})+e^{(b+1)k}-e^{k}\big\}}{(e^{k}-1)(e^{k}-z^{2})(z^{2}e^{k}-1)(z^{2(b+1)}-1)}e^{-bk/2}.
    \end{aligned}
\end{align}
\end{widetext}

Using Eq.~\eqref{eq:app_coefs} in the ansatz~\eqref{eq:app_ansatz}, we obtain the expression for the Fourier-transform of the total state density, which is valid on the entire domain of $k\in\mathbb{R}$. It can be compactly written as
\begin{align}
\label{eq:app_total-density_solution}
    \hat{\rho}^{{\rm tot}(b)}_{m}(k;\!z) = \frac{\mathcal{X}_m}{\mathcal{X}_b \mathcal{X}_{m-1}\mathcal{X}_{m+1}}\hat{\Xi}^{(\max{(m,b)})}_{\min{(m,b)}}(k;\!z),
\end{align}
where we have defined
{\allowdisplaybreaks
\begin{align}
    \hat{\Xi}^{(b)}_m(k;\!z)\!\equiv &\,e^{-(m+b+1)\tfrac{|k|}{2}}\!\big[\!\mathcal{X}_{b+1}(z)e^{|k|}\!-\!\mathcal{X}_{b-1}(z)\!\big]\notag\\
    &\times\!\sum_{j=0}^{m-1}\!\mathcal{X}_{j}(z)\mathcal{X}_{m-j-1}(z)e^{(m-j-1)|k|}=\notag\\
    =&\,e^{-(m+b+1)\tfrac{|k|}{2}}\!\sum_{j=0}^{m}\!\big[\!\mathcal{X}_{b+1}(z)\mathcal{X}_j(z)\mathcal{X}_{m-j-1}(z)\notag\\
    &-\!\mathcal{X}_{b-1}(z)\mathcal{X}_{j-1}(z)\mathcal{X}_{m-j}(z)\!\big]e^{(m-j)|k|}.
\end{align}
}
\noindent The latter function can be further simplified at half-filling $\mu=0$, where $\lim_{z\to 1}\mathcal{X}_m(z)=m+1$, and we obtain
\begin{align}
\label{eq:app_xi_half-filling}
    \hat{\Xi}^{(b)}_{m}\!(k;\!1)\!=&\,e^{-(m+b+1)\tfrac{|k|}{2}}\notag\\
    &\times\sum_{j=0}^{m}\left[m(2j\!+\!b\!+\!2)\!-\!2j(j\!+\!b\!+\!1)\right]e^{(m-j)|k|}.
\end{align}

\subsubsection{Drift velocity}

We now return to a chain of alternating spins $s_1$ and $s_2$. The single-magnon quasimomentum~\eqref{eq:quasimomentum} and quasienergy~\eqref{eq:quasienergy} in the integrable ratchet imply that the quasienergies and quasimomenta of quasiparticles (i.e., bound states of magnons) can be obtained from
\begin{align}
\begin{aligned}
    \boldsymbol{\varepsilon}&=\frac{1}{2}\left(\boldsymbol{p}^{(b_1)}_{+} -\boldsymbol{p}^{(b_2)}_{-}\right),\\ 
    \boldsymbol{p}&=\frac{1}{2}\left(\boldsymbol{p}^{(b_1)}_{+}+\boldsymbol{p}^{(b_2)}_{-}\right),
\end{aligned}
\end{align}
respectively, where the elements $p^{(b)}_m(\lambda_{\pm})$ of $\boldsymbol{p}^{(b)}_{\pm}$ denote the quasimomenta of $m$-strings in the homogeneous Heisenberg chain of spins $s=b/2$. Differentiating on $\lambda$, using the linear dressing equation~\eqref{eq:app_dressing}, in which $\mathbf{1}+\mathbf{K}\mathbf{n}$ does not depend on the spins $s_1$ and $s_2$, and invoking Eq.~\eqref{eq:app_dressed-momentum}, we now recognize 
\begin{align}
\begin{aligned}
\label{eq:app_dressed_derivatives}
    (\boldsymbol{\varepsilon}')^{\rm dr}&=\pi\left[\boldsymbol{\rho}^{{\rm tot}(b_1)}_{+}-\boldsymbol{\rho}^{{\rm tot}(b_2)}_{-}\right],\\
    (\boldsymbol{p}')^{\rm dr}&\equiv2\pi \boldsymbol{\rho}^{{\rm tot}}=\pi\left[\boldsymbol{\rho}^{{\rm tot}(b_1)}_{+}+\boldsymbol{\rho}^{{\rm tot}(b_2)}_{-}\right].
\end{aligned}
\end{align}
The lower labels $\pm$ again refer to the shift $\lambda_\pm=\lambda\pm\tau/2$ in the rapidity, i.e., in the continuous row index of a vector.
From here, the effective velocity~\eqref{eq:effective_velocity} of a quasiparticle in an alternating spin chain can be obtained, expressed in terms of the total state densities whose Fourier transforms are given in Eq.~\eqref{eq:app_total-density_solution}:
\begin{align}
\label{eq:app_eff-velocity-ratio}
    v^{\rm eff}_m(\lambda)=\frac{\rho_m^{{\rm tot}(b_1)}(\lambda_{+})-\rho_m^{{\rm tot}(b_2)}(\lambda_{-})}{\rho_m^{{\rm tot}(b_1)}(\lambda_{+})+\rho_m^{{\rm tot}(b_2)}(\lambda_{-})}.
\end{align}

In order to obtain the drift velocity~\eqref{eq:drift_velocity}, we now have to evaluate infinitely many integrals
\begin{align}
\begin{aligned}
\label{eq:app_integrals}
    \mathfrak{a}_m(\mu)&\!=\!(q_m^{\rm dr})^2\int {\rm d}\lambda\,\chi_{m}(\lambda)v^{\rm eff}_m(\lambda)\\
    &\!=\!(q_m^{\rm dr})^2\frac{\mathcal{X}_m^2\!-\!1}{2\mathcal{X}_m^4}\!\int {\rm d} \lambda\left[\rho^{{\rm tot}(b_{1})}_{m}(\lambda_{+})\!-\!\rho^{{\rm tot}(b_{2})}_{m}(\lambda_{-})\right],\\
    \mathfrak{b}_m(\mu)&\!=\!(q_m^{\rm dr})^2\int {\rm d}  \lambda\,\chi_{m}(\lambda)\\
    &\!=\!(q_m^{\rm dr})^2\frac{\mathcal{X}_m^2\!-\!1}{2\mathcal{X}_m^4}\!\int {\rm d} \lambda\left[\rho^{{\rm tot}(b_{1})}_{m}(\lambda_{+})\!+\!\rho^{{\rm tot}(b_{2})}_{m}(\lambda_{-})\right],
\end{aligned}
\end{align}
in terms of which $v_{\rm d}=(\sum_m \mathfrak{a}_m)/(\sum_m\mathfrak{b}_m)$.
Here, $\chi_m(\lambda)=\rho_m^{\rm tot}(\lambda)n_m (1-n_m)$ are mode susceptibilities, and the occupancy functions~\eqref{eq:app_occupancy} have been used. In passing to the second line in Eq.~\eqref{eq:app_integrals} we have used Eq.~\eqref{eq:app_eff-velocity-ratio} for the effective velocity and the total state density $\rho^{\rm tot}_m(\lambda)=[\rho^{{\rm tot}(b_1)}_m(\lambda_{+})+\rho^{{\rm tot}(b_2)}_m(\lambda_{-})]/2$, inferred from Eq.~\eqref{eq:app_dressed_derivatives}.

We will now assume that the result at half-filling ($\mu=0$) can be obtained by considering only the leading order in $\mu$ of all involved expressions, i.e., separately in the numerator and denominator of $v_{\rm d}$. In the leading order, the $\mathfrak{su}(2)$ characters $\mathcal{X}_m$ are independent of $\mu$, while the dressed magnetization is proportional to the chemical potential, $q_m^{\rm dr}=\tfrac{1}{6}\mu (m+1)^2+O(\mu^2)$. We then have
\begin{align}
\begin{aligned}
    \mathfrak{a}_m(\mu)&\!=\!\frac{\mu^2 m(m\!+\!2)}{72}\left[\mathfrak{I}_m^{(b_1)}(\tfrac{\tau}{2})\!-\!\mathfrak{I}_m^{(b_2)}(-\tfrac{\tau}{2})\right]\!+\!O(\mu^3),\\
    \mathfrak{b}_m(\mu)&\!=\!\frac{\mu^2 m(m\!+\!2)}{72}\left[\mathfrak{I}_m^{(b_1)}(\tfrac{\tau}{2})\!+\!\mathfrak{I}_m^{(b_2)}(-\tfrac{\tau}{2})\right]\!+\!O(\mu^3),
\end{aligned}
\end{align}
where we have defined $\mathfrak{I}_{m}^{(b)}(\nu)\equiv\int{\rm d} \lambda  \,\rho_{m}^{{\rm tot}(b)}(\lambda+\nu)$. We can evaluate them using the solutions of the Bethe-Yang equations~\eqref{eq:app_total-density_solution} at half-filling [i.e., together with Eq.~\eqref{eq:app_xi_half-filling}], obtaining
\begin{align}
    \mathfrak{I}_{m}^{(b)}(\nu)=\begin{cases}
        \frac{(m+1)^2}{3(b+1)} & m<b,\\[1em]
        \frac{b(b+2)(m+1)}{3m(m+2)} & m\ge b.
    \end{cases}
\end{align}
Notably, they are independent of $\nu$, so the drift velocity cannot depend on the parameter $\tau$ of the unitary gate. Splitting the sums in the drift velocity $v_{\rm d}=(\sum_m \mathfrak{a}_m)/(\sum_m\mathfrak{b}_m)$ into those over the intervals $m<\max(b_1,b_2)$ and $m\ge \max(b_1,b_2)$, we see that the latter will be divergent and the former negligible in comparison. Keeping only the terms with $m\ge \max(b_1,b_2)$ finally leads us to
\begin{align}
\label{eq:app_mu0-drift}
    v_{\rm d}(\mu=0)=\frac{b_1(b_1+2)-b_2(b_2+2)}{b_1(b_1+2)+b_2(b_2+2)},
\end{align}
which is equivalent to Eq.~\eqref{eq:casimir-velocity}.

\bibliographystyle{apsrev4-2}
\bibliography{references.bib}
\end{document}